\documentclass{aa}  

\usepackage{graphicx, natbib, txfonts}
\bibpunct{(}{)}{;}{a}{}{,} % to follow the A&A style

\usepackage{lscape}

\usepackage[colorlinks=true,citecolor=blue]{hyperref}

\begin{document}

   \title{Cascade adaptive optics with a second stage based on a\\  Zernike wavefront sensor for exoplanet observations}
   \subtitle{Experimental validation on the ESO/GHOST testbed}

   \author{M. N'Diaye\inst{\ref{inst1}}
   \and A. Vigan\inst{\ref{inst2}}
   \and B. Engler\inst{\ref{inst3}}
   \and M. Kasper\inst{\ref{inst3}}
   \and K. Dohlen\inst{\ref{inst2}}
   \and S. Leveratto\inst{\ref{inst3}}
   \and \\ J. Floriot\inst{\ref{inst2}}
   \and M. Marcos\inst{\ref{inst2}}
   \and C. Bailet\inst{\ref{inst1}}
   \and P. Bristow\inst{\ref{inst3}}
          }

   \institute{Université Côte d'Azur, Observatoire de la Côte d'Azur, CNRS, Laboratoire Lagrange, France\\
   \email{\href{mailto:mamadou.ndiaye@oca.eu}{mamadou.ndiaye@oca.eu}}
   \label{inst1}
   \and
   Aix Marseille Université, CNRS, CNES, LAM, Marseille, France
   \label{inst2}
   \and
   European Southern Observatory (ESO), Karl-Schwarzschild-Str. 2, 85748 Garching, Germany
   \label{inst3}
             }

   \date{Received XXX; accepted XXX}

% \abstract{}{}{}{}{} 
% 5 {} token are mandatory
 
  \abstract
    {Over the past decade, the high-contrast observation of disks and gas giant planets around nearby stars has been made possible with ground-based instruments using extreme adaptive optics (XAO).  These facilities produce images with a Strehl ratio higher than 90\% in the H band, in median observing conditions and high-flux regime. However, the correction leaves behind adaptive optics (AO) residuals, which impede studies of fainter or less massive exoplanets.}
   {Cascade AO systems with a fast second stage based on a Pyramid wavefront sensor (PWFS) have recently emerged as an appealing solution to reduce the atmospheric wavefront errors. Since these phase aberrations are expected to be small, they can also be accurately measured by a Zernike wavefront sensor (ZWFS), a well-known concept for its high sensitivity and moderate linear capture range. We propose an alternative second stage that relies on the ZWFS to correct for the AO residuals.}
   {We implemented the cascade AO with a ZWFS-based control loop on the ESO's GPU-based High-order adaptive OpticS Testbench (GHOST) to validate the scheme in monochromatic light. We emulated the XAO first stage in different observing conditions (wind speed, seeing) and determined the corresponding operation parameters (e.g., number of controlled modes, integrator gain, loop calibration) that lead to stable loop operation and good correction performance. Our strategy was assessed in terms of corrected wavefront errors and contrast gain in the images with a Lyot coronagraph to probe its efficiency.}
   {In median wind speed and seeing, our second-stage AO with a ZWFS and a basic integrator was able to reduce the atmospheric residuals by a factor of 6 and increase the wavefront error stability with a gain of 2 between open and closed loop. In the presence of non-common path aberrations, we also achieved a contrast gain of a factor of 2 in the coronagraphic images at short separations from the source, proving the ability of our scheme to work in cascade with an XAO loop. In addition, it may prove useful for imaging fainter or lighter close-in companions. In more challenging conditions, contrast improvements are also achieved by adjusting the control loop features.}
   {Our study validates the ZWFS-based second-stage AO loop as an effective solution to address small residuals left over from a single-stage XAO system for the coronagraphic observations of circumstellar environments. Our first in-lab demonstration paves the way for more advanced versions of our approach with different temporal control laws, non-linear reconstructors, and spectral widths. This would allow our approach to operate in high-contrast facilities on the current 8-10\,m class telescopes and Extremely Large Telescopes to observe exoplanets, all the way down to Earth analogs around M dwarfs.}


   \maketitle
%
%-------------------------------------------------------------------
\section{Introduction}
Unveiling the mechanisms of formation, evolution, and habitability of exoplanetary systems is one of the most exciting topics in modern astrophysics. The discovery of over 5500 exoplanets to date\footnote{e.g., see \url{https://exoplanet.eu/home/} or \url{https://exoplanetarchive.ipac.caltech.edu/}} with different detection methods has enriched our understanding of the demography and diversity of planetary systems, the morphology of circumstellar disks, and the internal structure and atmosphere of planets. High-contrast observation is a direct detection method that allows us to study the outer part of exoplanetary systems beyond the 5-10\,AU range, explore interactions between planets and disks, and characterize the atmospheric properties of exoplanets \citep[e.g.,][]{Currie2023,Chauvin2024}.

However, extracting the photons from exoplanetary companions is extremely challenging, given the contrast, namely, the flux ratio between a star and a planet, at short angular separation in the visible and near infrared \citep{Oppenheimer2009,Traub2010}. Nowadays, the foremost high-contrast facilities on the ground \citep[e.g., Gemini/GPI, Subaru/SCExAO, Keck/KPIC, VLT/SPHERE, Magellan/MagAO-X,][]{Macintosh2014,Jovanovic2015,Mawet2016,Beuzit2019,Males2022} or in space \citep[e.g., HST, JWST,][]{Debes2019,Carter2021,Hinkley2022} observe gas giant planets which are 10$^4$ to 10$^6$ times fainter than their stellar hosts at angular separations down to 200\,mas in emitted light. In the coming years, some of these exoplanet imagers will benefit from upgrades to image and spectrally analyze Jovian and Neptune-size planets, which are up to 10$^8$ times fainter than their host star at angular separations down to 100\,mas in thermal and reflected light \citep{Jovanovic2019,Ahn2021,Boccaletti2022,Chilcote2022,Males2022}. Over the next few decades, the Extremely Large Telescopes on the ground \citep[ELTs,][]{Kasper2021,Fitzgerald2022,Kautz2023} or observatories in space \citep[e.g., Roman Space Telescope, LUVOIR/HabEx mission concepts,][]{Bailey2023,LUVOIRreport,HabExreport} will benefit from high-contrast capabilities to study small gaseous and rocky planets that are 10$^8$ to 10$^{10}$ fainter than their parent stars at an angular separation of less than 20\,mas in reflected light.

High-contrast observations rely on different features such as extreme adaptive optics (XAO), wavefront sensing and control, stellar coronagraphy, observing strategies, and image post-processing methods. Instrumental and numerical innovations are required in all these aspects to gain up to four orders of magnitude in terms of contrast for the observation of these faint, close-in planetary-mass companions \citep{Galicher2024}. The associated technological leaps will lead to a refined understanding of the formation of Jovian and terrestrial planets, the survey of Earth analogs around Sun-like stars, and the tentative detection of biosignatures from nearby habitable zone exoplanets by the 2040s.

On the ground, the most advanced high-contrast facilities are built upon an XAO system to compensate for the effects of the atmospheric turbulence on the image of an observed star \citep{Guyon2018}. Usually based on a Shack-Hartmann wavefront sensor (SHWFS) to measure disturbances, these systems run at a typical speed of 1\,kHz to provide images with a Strehl ratio higher than 90\% in H-band in median conditions and in high-flux regime \citep[e.g.,][]{Perrin2003,Fusco2006,Beuzit2019}. Following the Maréchal approximation, the corresponding wavefront errors are of the order of 85\,nm Root Mean Square (RMS) or 0.05\,$\lambda$ at $\lambda$=1.65\,$\mu$m, where $\lambda$ denotes the wavelength of observation. 

While these XAO systems perform an excellent correction of the atmospheric wavefront perturbations \citep[e.g.,][]{Fusco2016}, they leave behind some adaptive optics (AO) residuals, which impede the observation of the substellar mass companions. Upgrading these key modules will pave the way for novel exoplanetary science capabilities, such as  access to young gas giant planets down to the snow line or the observation of a large number of red stars \citep{Boccaletti2020}. Observing fainter or closer planets to their star involves achieving image contrast larger than 10$^6$ down to 100\,mas and gaining sensitivity for red stars. 

To reduce AO residuals, the XAO systems require a more sensitive wavefront sensor (WFS) than SHWFS and a higher temporal bandwidth than 1\,kHz to achieve more accurate wavefront error measurements and faster corrections than the evolution of the atmospheric turbulence. Upgrading the existing XAO systems is an appealing solution but a risky option from the operational standpoint since the availability of the current functionalities of these exoplanet imagers is often requested by the community to preserve the existing high-contrast capabilities. To avoid modifying the initial XAO, a promising approach consists of using a cascade AO system \citep[e.g.,][]{Cerpa-Urra2022}. This concept consists of assisting the original XAO system with a subsequent AO stage by including its own deformable mirror (DM), WFS, and fast real-time computer (RTC) to decrease the atmospheric residuals further.

To accurately measure the residual wavefront errors left from the first stage, the second-stage AO loop relies on a WFS with high sensitivity and moderate capture range. Due to its characteristics, the Pyramid wavefront sensor \citep[PWFS,][]{Ragazzoni1996} appears as a relevant solution for the second stage to boost the wavefront correction of XAO systems. Such a PWFS-based approach is currently being considered or implemented in the near infrared in several high-contrast facilities on large ground-based telescopes \citep{Jovanovic2019,Ahn2021,Males2022,Boccaletti2022} to improve their AO performance and favor the observation of faint, low-mass companions down to 100\,mas from their host star.

The regime of expected atmospheric residuals also appears favorable to a second-stage AO loop with a Zernike wavefront sensor (ZWFS), a concept based on phase-contrast principle to convert wavefront errors into intensity variations \citep{Zernike1934}. Such a sensor is well-known for its moderate capture range and its high sensitivity \citep[e.g.,][]{Bloemhof2003,Dohlen2004,Guyon2005,Wallace2011,Jensen-Clem2012,N'Diaye2013a,Chambouleyron2021}. This concept has proven to be a promising solution for a wide range of astronomical applications, such as the measurement of non-common path aberrations (NCPA) in high-contrast instruments \citep[e.g., ZELDA on VLT/SPHERE,][]{N'Diaye2016,Vigan2019,Vigan2022}, the estimate of cophasing errors in segmented aperture telescopes \citep[][]{Dohlen2006,Surdej2010,Vigan2011,Janin-Potiron2017,Pourcelot2021,vanKooten2022,Salama2024},  measurement of fine low-order aberrations in space observatories with coronagraphic capabilities \citep[e.g.,][]{Shi2016,Pourcelot2022,Pourcelot2023}, or the picometric precision metrology for future space coronagraphs \citep[e.g.,][]{Ruane2020,Steeves2020}. 

In this paper, we investigate a ZWFS-based approach to run fast and accurate wavefront corrections on the image of an observed star after an XAO loop.  Our ZWFS-based AO loop is implemented on the GPU-based High-order adaptive OpticS Testbench (GHOST) at the ESO Headquarters in Garching (Germany) to demonstrate its efficiency in simulated median observing conditions. We consider different control loop aspects such as the number of corrected modes, the temporal loop gain, and the calibration to assess their impact on the wavefront correction. The ZWFS control loop is then explored through different observing conditions to derive the best functioning points. Finally, we perform some comparisons with the PWFS-based approach to derive the complementary working regimes for our ZWFS-based control loop. Our tests were run in autumn 2023. The whole study is carried out in the presence of a monochromatic light source to achieve a first experimental validation of our scheme.

\section{Experimental setup}
\subsection{GHOST testbed setup}
\begin{figure*}[!ht]
    \centering
    \includegraphics[width=\textwidth]{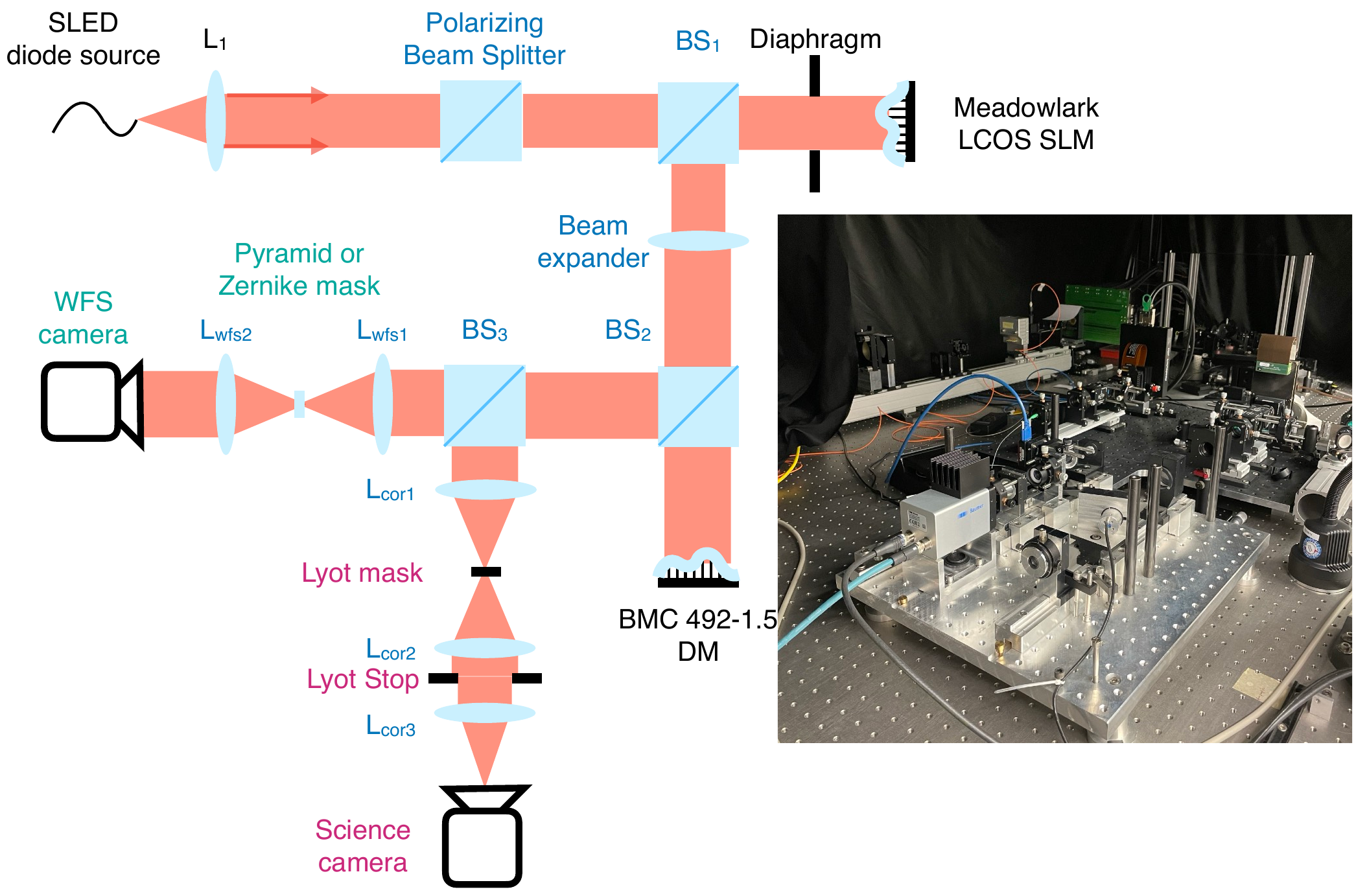}
    \caption{Presentation of the GHOST testbed. \textbf{Left}: Optical layout of the GHOST bench, with the following symbols: L: lens, BS: beam splitter, DM: deformable mirror, SLM: spatial light modulator. See text for more details. The modulation mirror used for the PWFS and located between BS3 and L$_{wfs1}$ is not represented in this scheme. \textbf{Right}: Picture of the testbed on March 8, 2023.
}
    \label{fig:ghost-scheme}
\end{figure*}

Figure \ref{fig:ghost-scheme} shows the optical layout and a picture of the GHOST testbed. We briefly recall the main features of the bench and highlight the parts that are relevant for our study.

GHOST is equipped with two different light sources: a light emitting diode (LED) and a super luminescent diode (SLED) emitting at 740nm and 770nm, respectively. Our work here relies on the use of the SLED source. Based on a single-mode fiber, a point-like source emits a light beam which is collimated by the achromatic lens L$_1$. The light then goes through a polarizing beam splitter cube, a standard beam splitter BS$_1$ and a 10\,mm-diameter diaphragm, leading to a polarized light beam that hits the Meadowlark liquid crystal on silicon (LCoS) spatial light modulator (SLM) in reflection to inject simulated XAO residual turbulence phase screens. 

The reflected light from the SLM travels back to BS$_1$ and is then resized with a beam expander to 6.7mm before hitting the second beam splitter BS$_2$. The transmitted light goes towards the 492-1.5 DM from Boston Micromachine Corporation (BMC) with an actuator pitch and stroke of 300\,$\mu$m and 1.5\,$\mu$m. The DM is used as a corrector for the second-stage AO to reduce the XAO residuals from the SLM. After reflection, the beam travels back to BS$_2$ and then faces the beam splitter BS$_3$, which  splits the light between the wavefront sensing and the science arms equally. 

The reflected light after BS$_3$ goes through a classical Lyot coronagraph (CLC) with a 4\,$\lambda$/D opaque focal plane mask (FPM) and a Lyot stop with a diameter of 0.84 times the pupil size $D$ \citep{Lyot1932,Vilas1987}, producing a coronagraphic image of the source on the Basler acA2040-90um science camera. The beam has a focal ratio F/23 and the camera pixel size is 5.5\,$\mu$m. The image plate scale is 3.22\,pixel per resolution element.

The transmitted light after BS$_3$ goes to the wavefront sensing arm with a lens that forms the source image on the sensor with a f/50 beam ratio. In this focal plane, a field stop can also be inserted to reduce aliasing effects in the WFS measurements \citep[e.g.,][]{Poyneer2004}. Field stops with diameters of 35, 49 and 63$\lambda/D$ are available to adjust as the function of the atmospheric conditions. A downstream lens forms the image of the relayed pupil onto a 10GigE Sony IMX CMOS camera to capture the sensor signal for the wavefront error measurements. The pupil is imaged with 36 pixels across its diameter. The standard GHOST setup uses the PWFS to estimate the residual atmospheric aberrations. The wavefront sensing arm also includes an additional PI SL-325 modulation mirror to enable beam modulation on the PWFS. This device is not represented in the scheme. In this work, we replaced the Pyramid with a Zernike phase-shifting mask.

\subsection{Zernike wavefront sensor}
\subsubsection{Principle}
\begin{figure}[!ht]
    \centering
    \includegraphics[width=\linewidth]{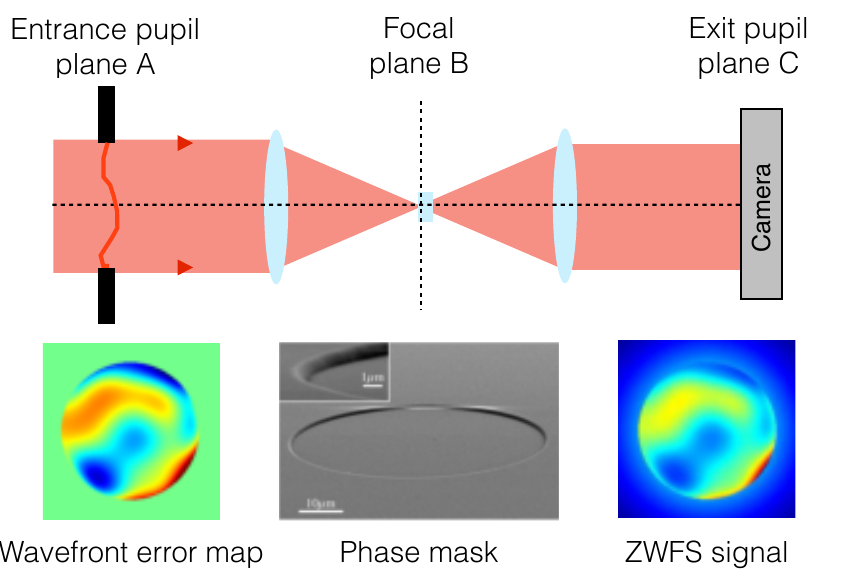}
    \caption{Schematic diagram of the ZWFS analysis with the wavefront errors in the entrance pupil to be estimated, a phase-shifting mask centered on the on-axis stellar point source at the focus of the telescope aperture and the intensity measurement in the relayed pupil plane. In the small aberration regime ($\varphi \ll$ 1\,rad), a linear reconstruction of the aberrations is performed from the recorded intensity with a nanometric accuracy.}
    \label{fig:zelda-scheme}
\end{figure}

We briefly recall the principle of the ZWFS, see its optical layout in Fig.~\ref{fig:zelda-scheme}. The light beam from a point-like source propagates through the sensor, first going through the telescope pupil in plane A in which phase aberrations $\varphi$ are present. The source image is then formed in the following focal plane B, where a mask with a phase shift of $\theta$ and a diameter, $d$, of about one resolution element is located. The light going through and surrounding the mask dot generates interferences in the relayed pupil plane C, producing an intensity pattern that is directly related to $\varphi$. In recent years, the formalism of the ZWFS has been largely described for applications in astronomy and further details can be found in the literature \citep[e.g.,][]{N'Diaye2013a,Ruane2020,Chambouleyron2021}. Following \citet{N'Diaye2013a}, the relayed pupil intensity $I_C$ is expressed as a function of the phase $\varphi$ for a given pixel in the pupil with
\begin{equation}
    I_C = P^2 + 2b^2\,(1-\cos{\theta})+2Pb\ [ \sin\varphi\sin\theta-\cos\varphi(1-\cos\theta)\ ]\,.
    \label{eq:zwfs_signal}
\end{equation}
where $P$ and $b$ represent the amplitude pupil function and the amplitude diffracted by the focal plane phase mask. 
In the absence of aberrations ($\varphi=0$), the sensor response is non-null and its intensity, $I_C^0$, is expressed as
\begin{equation}
    I_C^0 = P^2 + 2b^2\,(1-\cos{\theta})-2Pb\ [ (1-\cos\theta)\ ]\,.
    \label{eq:zwfs_signal0}
\end{equation} 
In the small aberration regime ($\varphi \ll$ 1\,rad), we refer to the first-order Taylor expansion of the terms with $\varphi$ to derive a linear relation from Eq.~ (\ref{eq:zwfs_signal}) between the intensity variations in the relayed pupil and the wavefront errors in the entrance pupil. Such an approach favors a simple and fast retrieval of small aberrations. Assuming that $P$, $b$, and $\sin\theta$ are not equal to zero, the phase is then derived from $I_C$ with
\begin{equation}
    \varphi = \frac{I_C-P^2+2b(1-\cos\theta)(P-b)}{2Pb\sin\theta}\,.
    \label{eq:zwfs_phase}
\end{equation}
The phase can also be expressed by accounting for $I_C^0$ from Eq.~(\ref{eq:zwfs_signal0}) with
\begin{equation}
    \varphi = \frac{I_C-I_C^{0}}{2Pb\sin\theta}\, .
\end{equation}
These expressions are valid for the phase aberrations which correspond to a $\pm 0.05\,\lambda$ range for the optical path differences. In this linear regime, the phase aberrations can alternatively be reconstructed with the ZWFS by relying on a command matrix built from an interaction matrix, following the standard approach in AO. In our approach, the phase retrieval method with ZWFS is based on the use of interaction and command matrices. 

\subsubsection{GHOST mask prototype}
The GHOST Zernike mask is a fused silica substrate in which a dot is machined by SILIOS technologies with photolithographic reactive ion-etching \citep{N'Diaye2010} to achieve a hole with a diameter and depth of 40.5 and 0.423\,$\mu$m. At $\lambda$=770\,nm and with f/50 beam ratio, the mask has a relative size of 1.05$\lambda/D$ and introduces a $\pi$/2 phase shift on a fraction of the source image at the focal plane in the wavefront sensing path on GHOST. 

The metrology of the prototype was carried out at Laboratoire d'Astrophysique de Marseille with a Wyko NT9100 interference microscope in vertical scanning mode, allowing us to confirm that the measured mask diameter and depth are within 1\% of the specifications that were given to the manufacturer. For our tests, the mask is mounted on xy-translation stages to adjust its position with respect to the source image.  

\subsection{Experimental protocol}\label{sec:protocol}
In the following, we describe the protocol with the four main steps of our experiments. The first step consists of injecting phase screens of atmospheric turbulence residuals on the SLM with a frame rate up to 400\,Hz. The phase screens have been derived by a numerical simulation using the Object-Oriented Python and Adaptive Optics (OOPAO) simulation tool \citep{Heritier2023}. Here, the first stage AO was correcting for 800 Karhunen–Loève (KL) modes using a PWFS and a 40$\times$40 actuator DM using a sixth-magnitude star and various wind speed and seeing combinations. The simulated atmospheric turbulence was updated at 2\,kHz, but the control loop ran at half that speed by using two subsequent averaged WFS images to update the DM at 1\,kHz. The residual phase screens from these simulations at 2\,kHz were saved in data cubes to be replayed by the SLM at a speed of 350\,Hz. These residuals provide a realistic representation of the atmospheric wavefront errors left after correction by an XAO instrument such as VLT/SPHERE.

The second step deals with the calibration of the ZWFS-based control loop. After setting up the SLM to a flat position, an interaction matrix is first built by sending Hadamard modes \citep{Kasper2004} to the BMC-492 with 24 actuators across the pupil diameter. The sensor response is recorded on the camera with 36 pixels across the pupil diameter. We then determine the command matrix by calculating the KL modal interaction matrix and computing the pseudo-inverse of this modal interaction matrix for the desired number of modes. This procedure is repeated twice in a row. The first interaction matrix is measured in the presence of aberrations on the Zernike mask. The resulting control matrix is used to close the loop, setting the WFS reference signals to zero. This results in a flat wavefront on the Zernike mask as bench aberrations are compensated by the DM. The corresponding DM position is then used for a second interaction matrix calibration with the ZWFS working around its zero-aberration position. This procedure leads to a better calibration of the ZWFS and improved closed loop robustness. We evaluate the effects of both calibration steps in Sect. \ref{subsec:calibration}.  

The third step is related to the temporal control of the AO residuals. The second-stage AO loop is driven with the COSMIC platform \citep{Ferreira2022}, a GPU-based RTC with 2 CPUs and 112 cores in total, and 2 Titan RTX GPUs. It uses a standard AO pipeline, which shows a delay of 110\,$\mu$s between the last pixel received from the WFS and the voltage sent to the DM. The COSMIC Graphics User Interface (GUI) allows us to remotely control the DM, the WFS camera, and the modulation mirror for tests with the PWFS. For the experiments with the ZWFS control loop, no modulation is required and the modulation mirror is left unused in a static position. An additional computer with remote access allows us to drive the source illumination, the SLM, the camera in the science arm, and finally the viewer on the wavefront sensing camera with the standalone software. All the functions are remotely accessible independently, and an integrated version of all the software is currently under development.

For the temporal controllers, classical leaky integrator control and reinforcement learning schemes \citep[e.g.,][]{Nousiainen2024} are available with the GHOST interface to explore different control strategies. For the current study with the ZWFS, we limit our temporal control to a classical integrator with a gain and leak to mimic standard AO control with the second-stage AO. Further tests with more elaborate real-time control laws will be considered in the near future. Unless otherwise stated, the ZWFS-based control loop runs at the same speed as the SLM (350\,Hz) to simulate a second stage with a speed that is twice the update rate of the first stage.

Finally, the fourth step corresponds to the data acquisition in open and closed loop for the second-stage AO for performance assessment. In the following, we present our measurements of the residual wavefront errors from the WFS path to estimate the AO correction. The residual WFS measurement is given in normalized DM commands [V] after multiplication by the control matrix (0-200 V are commanded by values ranging between 0 and 1). We also acquired data on the science arm with or without coronagraph to estimate the contrast difference in the normalized coronagraphic images between the open and closed loop situations. The images were sampled with 3.22\,pixel per resolution element. 

\section{Lab demonstration in median conditions}
\subsection{Assumptions}
We ran a first experiment with the ZWFS-based second-stage AO loop using standard observing conditions to provide a proof of concept. The SLM injects phase screens corresponding to XAO residuals based on VLT/SPHERE characteristics under median observing conditions: a 10\,m.s$^{-1}$ wind speed and 0.7" seeing. The size of the field stop was set to 35\,$\lambda$/D.

\subsection{Correction of the AO residuals}
\begin{figure}[!ht]
    \centering
    \includegraphics[width=\linewidth]{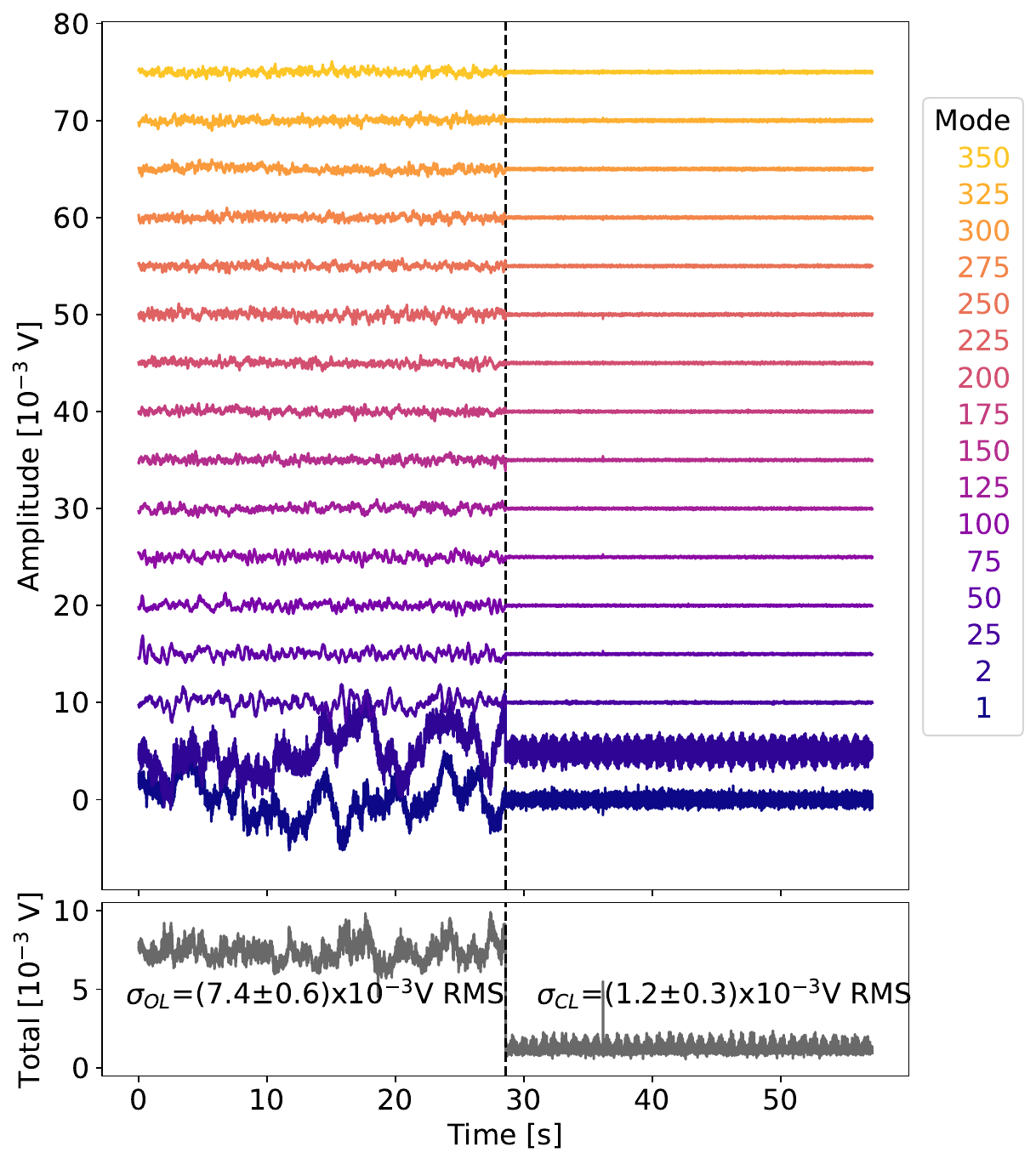}
    \caption{Temporal evolution of the wavefront errors. \textbf{Top}: Evolution for a few single modes. For each mode, the wavefront errors have been artificially shifted along the vertical axis to enhance readability: all of them actually oscillate around zero in closed loop. \textbf{Bottom}: Evolution for the total amount. In both panels, the dashed vertical line delimits the transition from open (left) to closed loop (right).}
    \label{fig:wfe_median_timeseries}
\end{figure}

After the generation of the command matrix, we operated the ZWFS control loop with a classical integrator for the temporal control law with a gain of 0.8 and a leak of 0.99. In closed loop, the number of corrected KL modes is set to 350. 
Figure \ref{fig:wfe_median_timeseries} top plot shows the temporal evolution of the wavefront errors for a subset of modes before and after closing the ZWFS control loop, open loop (OL) and closed loop (CL) respectively. A reduction of the temporal standard deviation is observed for all the modes and is mainly noticeable for the low-order modes, in particular the modes 1 and 2, which can be associated with pointing errors. 

\begin{figure}[!ht]
    \centering
    \includegraphics[width=\linewidth]{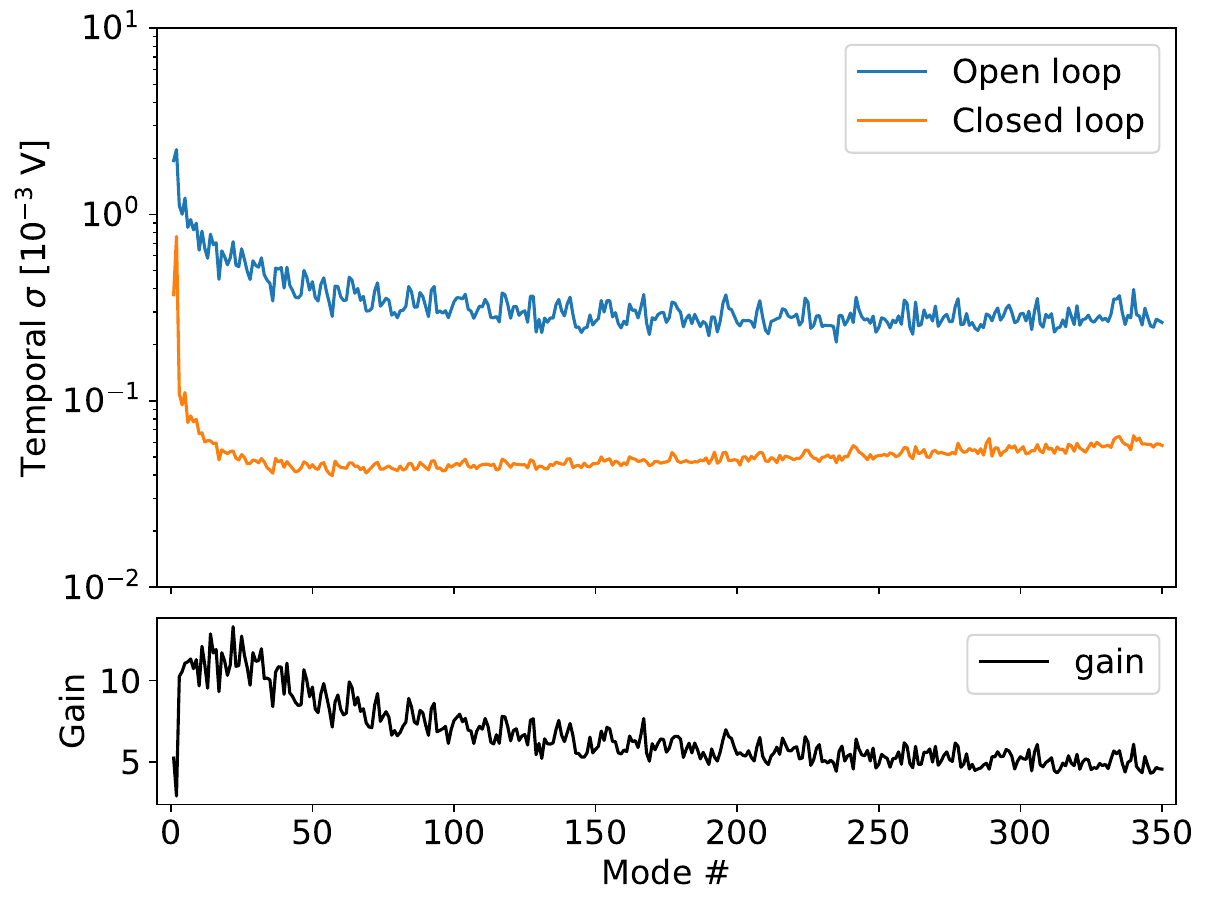}
    \caption{Temporal standard deviation of the wavefront errors for different modes with the ZWFS-based wavefront control. \textbf{Top}: Results in open (blue) and in closed loop (orange). \textbf{Bottom}: Reduction gain in wavefront error provided by the closed loop for each mode.}
    \label{fig:wfe_median_all_gains}
\end{figure}

Figure \ref{fig:wfe_median_all_gains} shows the temporal standard deviation $\sigma$ of the wavefront errors for each mode before and after closing the loop in top plot and the effective gain in the bottom plot. Our ZWFS-based second stage AO loop reduces the wavefront errors by at least a factor of 5 for all the modes and shows a gain up to about 13 for the low-order modes. In closed loop, the wavefront errors slightly increase with the highest modes and studies are on going to understand this point further. The observed effect could find its origin in the small difference in signal sampling between the ZWFS and the DM, respectively 36 pixels and 24 actuators across the pupil diameter. In this configuration, the transfer function of the sensor pixel introduces some sensitivity loss in the control loop for the modes with large spatial frequency content and additional calibration may be required to address this undesired feature. Still, this effect remains negligible in comparison with the large wavefront error reduction between the open and closed loop operations.

The wavefront error reduction is also visible when we consider the quadratic sum of all the modes to analyze the total amount of aberrations in open and closed loop (see Fig.~\ref{fig:wfe_median_timeseries}, bottom plot). We estimate the temporal mean of the RMS wavefront error $\sigma_{OL}$ and $\sigma_{CL}$ of (7.4$\pm$0.6)$\times 10^{-3}$ and (1.2$\pm$0.3)$\times 10^{-3}$V in open and closed loop, showing an overall reduction of the wavefront errors by a factor of 6 and a decrease in the temporal wavefront error dispersion by a factor of 2, confirming the ability of the ZWFS control loop to reduce the AO residuals and increase the wavefront error stability. In closed loop, we also observe a $\sim$1\,s cyclic behavior of the total wavefront errors which is also clearly present in Mode 2. This $\sim$1\,Hz appears to be a beat frequency since zooming in, there seems to be a very high frequency pattern, close to the Nyquist frequency. Actually, our analysis shows a good fit between the Mode 2 data and a sine wave of period 1.6\,ms (625\,Hz), multiplied by a squared cosine of period 1.6 seconds, producing the beating period of 0.8s (1.2\,Hz). This would correspond to the sum of two sine waves separated in frequency by 2.4\,Hz. We are currently investigating the origin of this effect, which could be related to vibrations in the bench or undesired effects in the temporal control. 

\begin{figure}
    \centering
    \includegraphics[width=\linewidth]{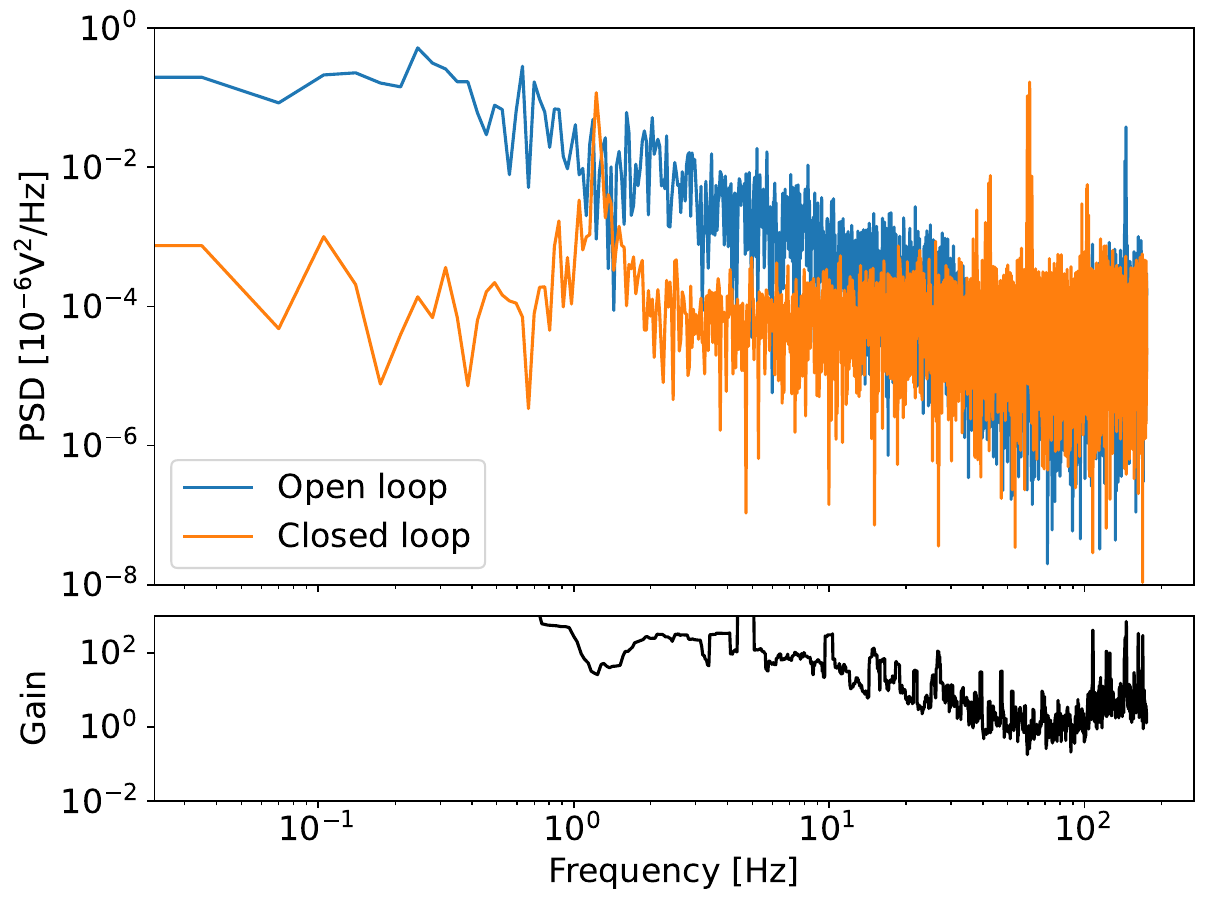}
    \caption{Temporal PSD of the wavefront errors with the ZWFS-based wavefront control and displayed in Fig. \ref{fig:wfe_median_timeseries} (bottom plot). \textbf{Top}: Results in open (blue) and closed loop (orange). \textbf{Bottom}: Reduction gain in PSD provided by the closed loop. The curve is a smoothed version of the original values for which a sliding average is applied for the sake of clarity.}
    \label{fig:wfe_mdian_psd}
\end{figure}

We also explored the temporal behavior of our second-stage AO loop by representing the temporal power spectral density (PSD) of the residual aberrations in open and closed loops, see Fig.~\ref{fig:wfe_mdian_psd}. Our loop enabled us to reduce aberrations by at least two orders of magnitude for wavefront errors with temporal frequencies up to 10\,Hz and larger than 1 for phase aberrations with temporal frequencies up to about 30\,Hz. The $\sim$1\,Hz peak observed in closed loop corresponds to the cyclic behavior that is reported in Fig.~\ref{fig:wfe_median_timeseries} (bottom plot). At higher temporal frequencies, the PSD shows slightly higher residuals, underlining some artifacts introduced by our control loop. However, the wavefront errors are small enough at these temporal frequencies to have no impact on the overall wavefront correction. 

\subsection{Contrast improvement in coronagraphic images}
We go on to study the impact of the AO residuals correction into contrast improvement in the coronagraphic image of the source. Our experiment was performed with a CLC and in the presence of NCPA between the sensing path and the science arm, which are left uncorrected in this experiment; thus, this limits the ultimate achievable contrast with our control loop.

\begin{figure}[!ht]
    \centering
    \includegraphics[width=\linewidth]{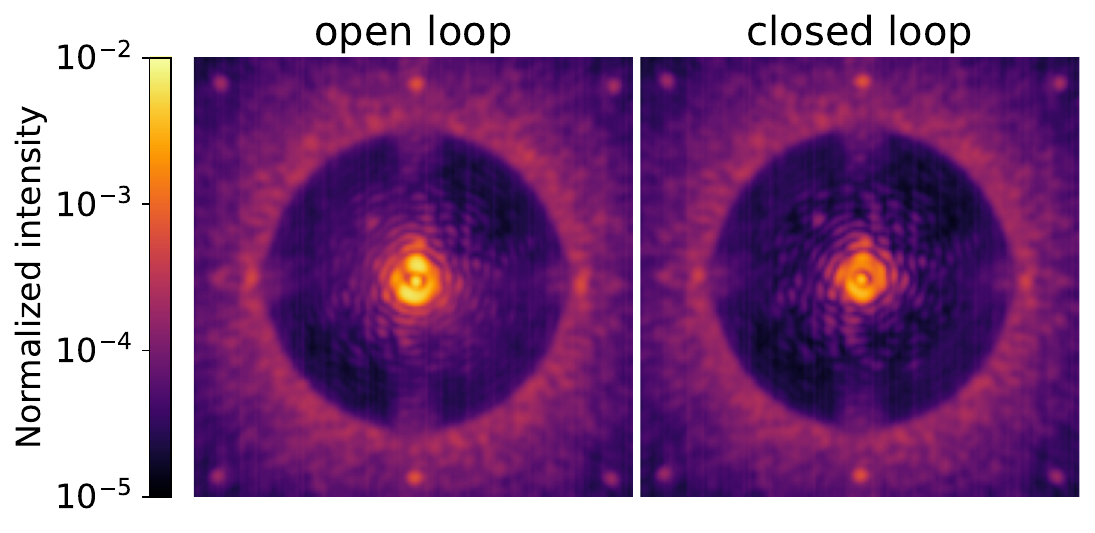}
    \includegraphics[width=\linewidth]{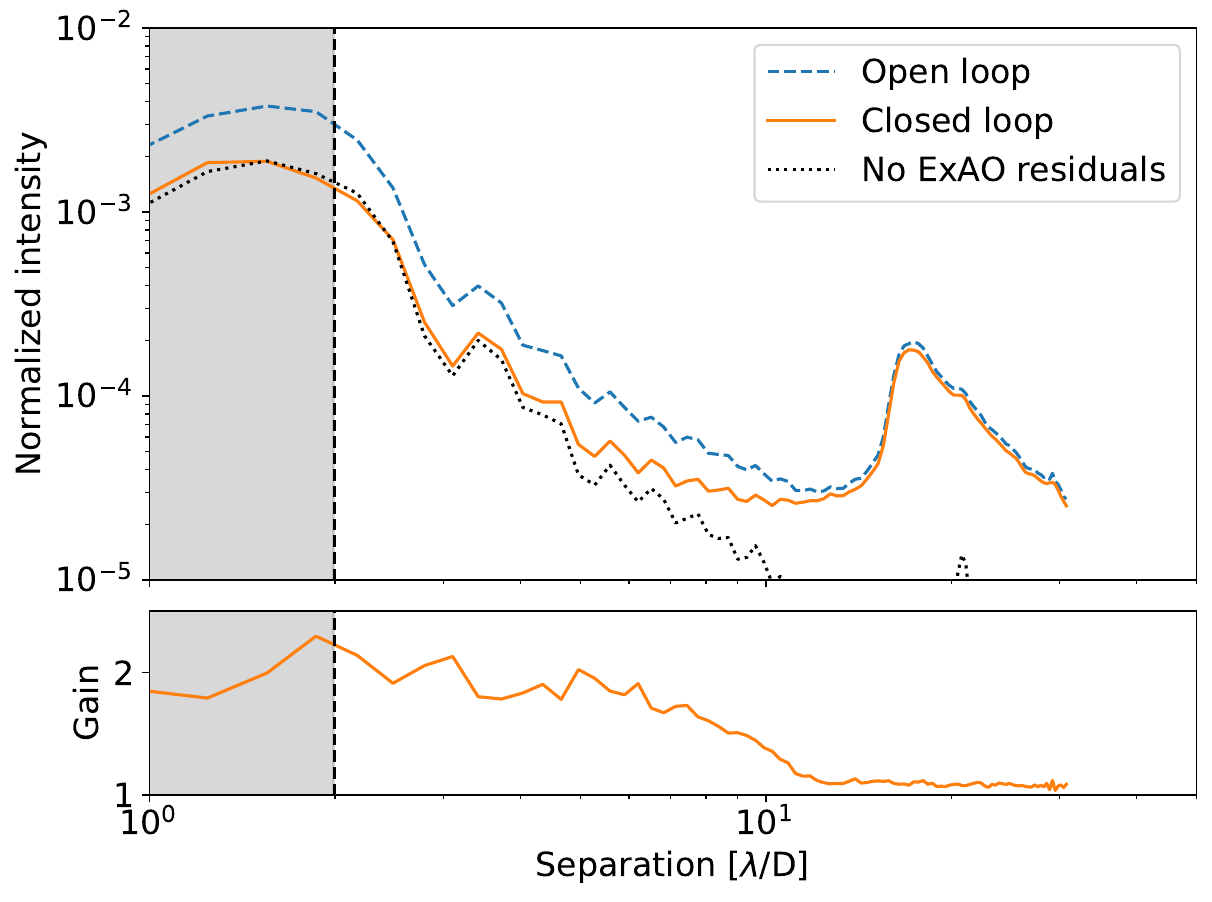}
    \caption{Contrast in the coronagraphic images with the ZWFS-based wavefront control. \textbf{Top}: Left and right, frames in log scale before and after closing the ZWFS control loop. The AO residuals based on VLT/SPHERE characteristics with a 6-mag natural guide star and median observing conditions with 10\,m.s$^{-1}$ wind speed and 0.7" seeing. In closed loop, the ZWFS-based second-stage AO controls 350 KL modes and runs with an integrator using a gain of 0.8. \textbf{Middle}: Normalized azimuthal averaged intensity profile of the coronagraphic images produced in open loop (blue dashed line), in closed loop (orange solid line) and in the absence of XAO residuals (black dotted line) as a function of the angular separation. The grey area with dashed lined delimits the projected FPM size. \textbf{Bottom}: Contrast gain provided by the ZWFS-based wavefront control between the open and closed loop operations.}
    \label{fig:cor_median_cond}
\end{figure}

Figure \ref{fig:cor_median_cond} top plot shows the recorded coronagraphic images in the science camera in open and closed loops. The anatomy of such images has been described in the literature \citep[e.g.,][]{Guyon2018,Cantalloube2019} and here, we briefly recall the origin of the main features. The images exhibit a large high-contrast region whose radius corresponds to the spatial frequency cutoff of the first stage XAO wavefront correction. 
Located at the edge of the images, some bright speckles outside the control region are produced by the DM actuator print-through. Compared to the open loop image, the closed-loop image shows the formation of an additional high-contrast region inside the XAO control region, at a distance from the source corresponding to a spatial frequency of about 11\,cycle/pupil (c/pup). This value is consistent with the expected frequency cutoff of the second-stage AO loop with the ZWFS for the correction of 350 KL modes ($\sqrt{350/\pi}$), confirming the effective correction performed by our control loop. 

Figure \ref{fig:cor_median_cond} (middle and bottom plots) shows the azimuthally averaged intensity profiles of the coronagraphic images in open and closed loops and the estimated contrast gain between them. We observe a contrast improvement with a gain by a factor of up to 2 in the ZWFS-based controlled region ranging from 2 to 11\,$\lambda/D$, showing the ability of our control loop to improve the contrast in the coronagraphic image. In the middle plot, we also represent the azimuthally averaged intensity profile of the coronagraphic in the absence of XAO residuals on the SLM, representing the ultimate contrast floor in our experiment. Our control loop manages to achieve this threshold at the separations out to 8\,$\lambda/D$. This contrast floor in our images finds several explanations, the main limitations in our experiment are most likely the limited contrast provided by the CLC and the presence of the NCPA between the wavefront sensing path and the science arm which lead to quasi-static speckles that alter the image contrast. Based on the measurements with the Fast \& Furious phase diversity algorithm \citep{Keller2012,Korkiakoski2014,Wilby2018,Bos2020} on the science camera, we estimate a total amount of NCPA of 20\,nm RMS for the first 36 Zernike modes after piston.

Despite these limitations, the overall results demonstrate the ability of the ZWFS control loop to reduce the AO residuals left from the first AO stage and increase the contrast in the coronagraphic image of an observed star to observe planetary companions at the shortest angular separations. %\kdo{KDo: any idea of the gain expected with a better coronagraph and corrected NCPA?}

In Appendix \ref{sec:appendix}, we analyze the impact of different control loop features in our approach, such as the number of corrected KL modes, the control loop gain, the size of the field stop, and the control loop calibration. These parameters constitute interesting degrees of freedom for our control loop to stabilize or possibly enhance the gain in coronagraphic images in different configurations. 

These promising results for the ZWFS control loop are obtained in median observing conditions. The behavior of our control loop needs to be further investigated to establish its performance characteristics in different observing conditions. The parameters and results are summarized in Appendix \ref{sec:summary}.

\section{ZWFS control loop through different observing conditions}
We here explore our second stage AO loop for several observing conditions to determine its functioning points in different environments. In nominal conditions, the SLM injects phase maps corresponding to AO residuals with a VLT/SPHERE-like instrument with a wind speed of 10\,m.s$^{-1}$ and 0.7" seeing. The field stop is set with a diameter of 35\,$\lambda/D$ for our experiments. Our second-stage AO loop is run to control 350 KL modes. The source flux corresponds to an intensity of 25\,$\mu$A with the SLED.

\subsection{Wind speed}
In XAO instruments, the observation of planetary companions or disks around an imaged star can be altered by the presence of a bright stellar veil called wind-driven halo \citep{Cantalloube2018,Cantalloube2020}. This harmful feature is related to bad observing conditions for which the AO loop runs slower than the atmospheric turbulence evolution. One of the main drivers is the wind speed which varies in amplitude and direction with altitude and time. Under these observing conditions, a single-stage XAO sees its atmospheric aberration compensation limited by the large temporal error between the wavefront error measurement and correction in the AO control loop. The so-called AO servo-lag error can be reduced at (i) the software level, for instance, by using predictive control strategies or PSF reconstruction for post-processing or at (ii) the hardware level with the use of a faster AO control loop. For the latter, a second-stage AO with a faster control loop constitutes an attractive solution to reduce the temporal error and address the residuals left from the first XAO stage. This approach is here assessed with our ZWFS-based control loop.

\begin{figure}[!ht]
    \centering
    \includegraphics[width=\linewidth]{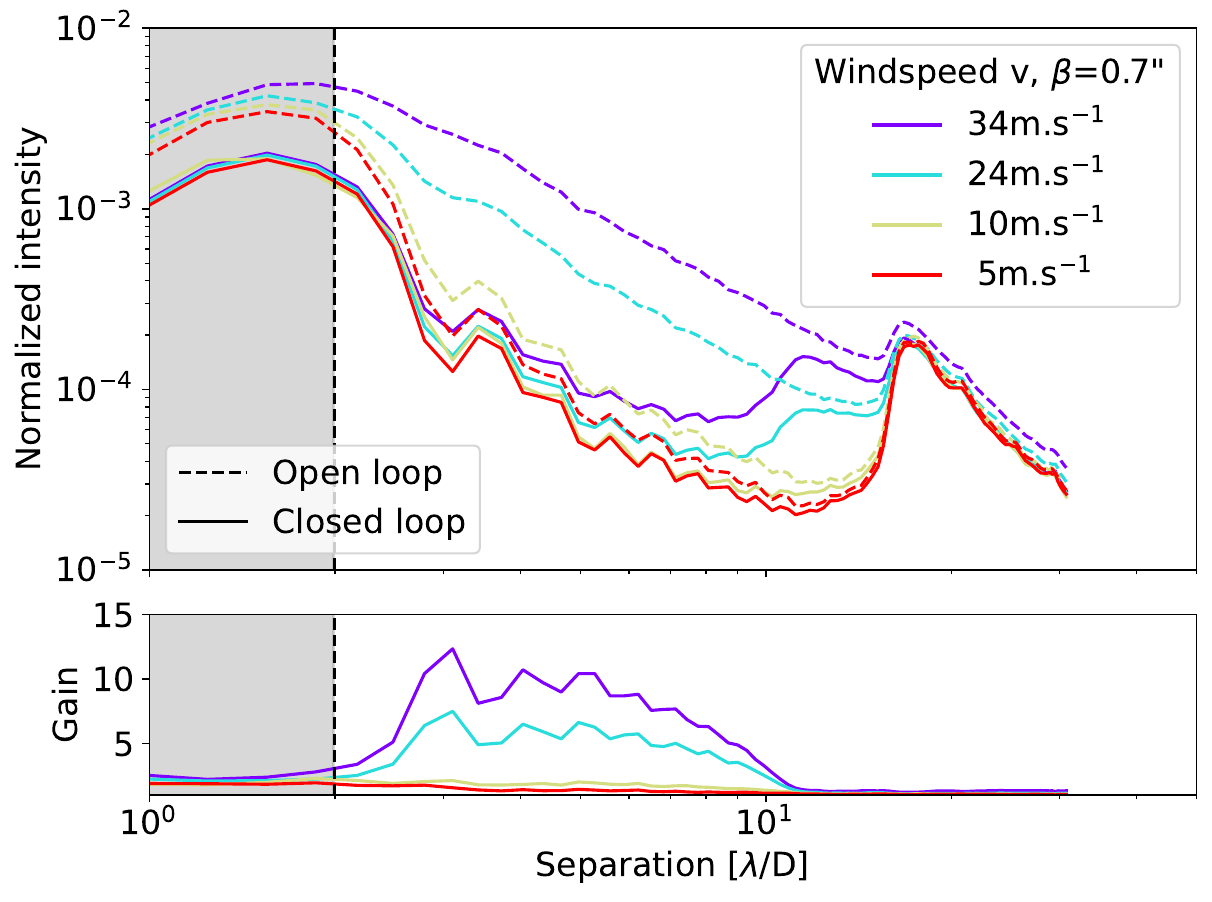}
    \caption{Contrast in the coronagraphic images with the ZWFS control loop for different wind speed conditions. \textbf{Top}: Normalized azimuthal averaged intensity profile of the coronagraphic images produced with the ZWFS wavefront control in open loop (dashed line) and closed loop (solid line) as a function of the angular separation for different windspeed conditions. The grey area with dashed line delimits the projected FPM size. The AO residuals are based on the VLT/SPHERE characteristics with a 6-mag natural guide star and median observing conditions with 0.7" seeing. In closed loop, the ZWFS-based second-stage AO controls 350 KL modes and runs with an integrator using a loop gain of 0.8. \textbf{Bottom}: : Contrast gain provided by the ZWFS-based wavefront control between the open and closed loop operations for different wind speeds.}
    \label{fig:cor_prf_windspeed}
\end{figure}

Figure \ref{fig:cor_prf_windspeed} represents the intensity profiles of the coronagraphic images in open and closed loop for different wind speeds, with a median seeing $\beta$ of 0.7". The prevailing effect here is the evolution speed of the wavefront errors. Our second-stage AO loop runs at a frequency which is twice faster than the first-stage XAO loop, enabling correction of the wavefront errors with a temporal frequency content that are unreachable with the first stage. The presence of the second-stage AO loop allows us to achieve contrast gain larger than 10 between the open and closed loops configuration for $v$ larger than 10\,m.s$^{-1}$. Theoretically, the temporal error goes with the correction bandwidth at the power of -5/3. This contrast gain with a twice larger frame rate comes from the fact that the first stage simulations assume a two-frame delay, while the second stage has a delay that is only slightly above one frame. So the correction bandwidth of the second stage is more than twice larger than the one of the first stage, leading to a reduction of the temporal error by a factor of $2^{5/3}$ and a contrast gain of $4^{5/3}$ ($\sim 10$). This outcome confirms the importance of having a fast loop correction in the presence of strong wind. 

In closed loop, a contrast floor is reached with the intensity profile of the coronagraphic image at the slowest wind speed with $v$ of 5\,m.s$^{-1}$. At slow wind speed, the contrast threshold is most likely dominated by the coronagraphic residuals and the NCPA at short separations in the ZWFS-based controlled region. The large wind speed cases clearly show the correction radius of the second-stage AO loop at 11\,$\lambda$/D while the controlled region is filled up by the speckles due to the NCPA and the diffraction residuals from the coronagraph for the slower wind speed cases.

We repeated a similar experiment with different $v$ (5 and 10\,m.s$^{-1}$) and $\beta$ of 1.0" and observed the same trends in terms of contrast gain and threshold level, although at a natural less deep contrast. At a higher wind speed ($v$=34\,m.s$^{-1}$), the control loop breaks down and no more correction is applied. The first stage AO residuals are too large for the ZWFS to remain in the linear range. Sign changes in the reconstructed wavefront errors will result in rapid loop divergence. The stability of the control loop is recovered by reducing $N_{modes}$ down to 250 modes. 

Our scheme runs efficiently in different wind speed conditions and can be maintained stable by adjusting $N_{modes}$ accordingly with the atmospheric turbulence conditions with $v$ and $\beta$. Compared with the first stage standalone, our approach allows us to gain contrast in all situations, providing access to fainter low-mass companions at closer separations from their host star. 

\subsection{Seeing}
Together with wind speed, another key feature of the atmospheric turbulence conditions is the seeing. This term denotes the PSF full width at half maximum, which increases with the spatio-temporal fluctuations of the atmospheric perturbation, leading to large wavefront errors to compensate for. At large seeings, these wavefront errors prove too large to be fully corrected for by a single XAO, limiting the contrast in the coronagraphic images of an observed star. To overcome the contrast threshold due to the atmospheric turbulence, addressing the atmospheric residuals with a second-stage AO is again a promising solution to enlarge the science return of high-contrast facilities over a wider range of seeing conditions. Here, we investigate the capabilities of our ZWFS-based control loop for different seeing conditions.

\begin{figure}[!ht]
    \centering
    \includegraphics[width=\linewidth]{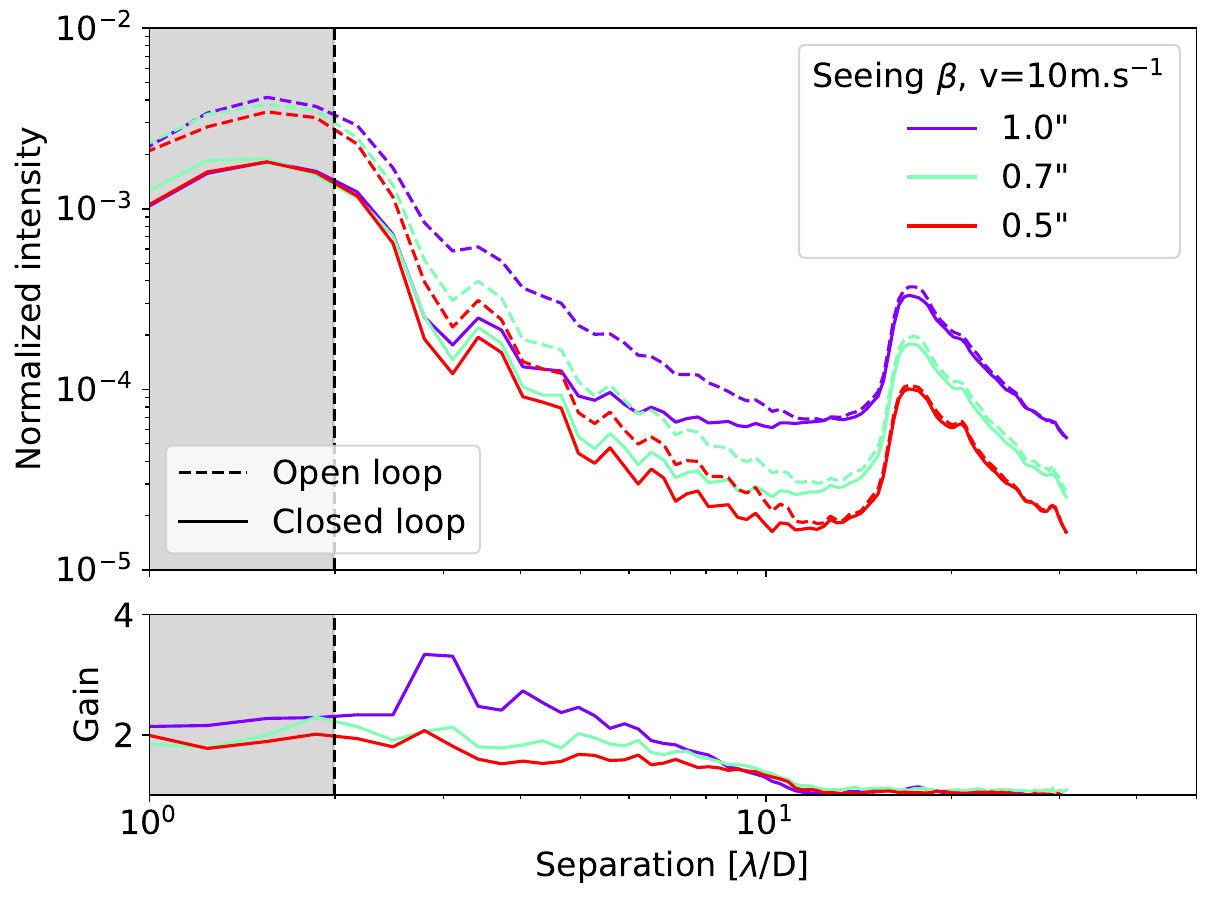}
    \caption{Contrast in the coronagraphic images with the ZWFS control loop for different seeing conditions. \textbf{Top}: Normalized azimuthal averaged intensity profile of the coronagraphic images produced with the ZWFS wavefront control in open loop (dashed line) and closed loop (solid line) as a function of the angular separation for different seeings. The grey area with dashed line delimits the projected FPM size. The AO residuals are based on the VLT/SPHERE characteristics with a 6-mag natural guide star and median observing conditions with 10\,m.s$^{-1}$ windspeed. In closed loop, the ZWFS-based second-stage AO controls 350 KL modes and runs with an integrator using a loop gain of 0.8. \textbf{Bottom}: Contrast gain provided by the ZWFS-based wavefront control between the open and closed loop operations for different seeings.}
    \label{fig:cor_prf_seeing}
\end{figure}

Figure \ref{fig:cor_prf_seeing} shows the intensity profiles of the coronagraphic images in open and closed loops for different seeing conditions with a wind speed of 10\,m.s$^{-1}$. The dominant effect is the amount of aberrations which is related to $\beta$, which denotes the seeing full width at half maximum and is expressed in arcsec. In open loop, the impact of the seeing is already observable with the intensity level of XAO control radius raising as the $\beta$ increases. In closed loop, the ZWFS control loop allows us to increase the contrast in the second stage controlled region in all the seeing cases. 

Our scheme achieves the largest contrast gain for the largest values of $\beta$ corresponding to conditions in which the wavefront error residuals from the first AO loop are the more important. For instance, a gain by a factor of up to 3.3 is observed at a separation of 3\,$\lambda/D$ for $\beta$ of 1.0". These results emphasize the benefit of having a second-stage AO loop with a faster correction to compensate for the XAO residuals when the observing conditions are less favorable, thereby extending the science return for exoplanet imaging. The ZWFS-based control loop still offers a contrast gain up to 2 at 3\,$\lambda/D$ for the best seeing with $\beta$=0.5", enabling access to the faintest planets in excellent observing conditions. 

In our experiments, some instabilities in the control loop are observed for $\beta$ of 1.0", leading to a loop break. These effects can be removed by simply reducing $N_{modes}$ from 350 to 300. Our ZWFS control loop runs efficiently in different seeings by simply adjusting the number of corrected KL modes.

\subsection{Source flux}
We studied the impact of the source flux on the wavefront error correction with our control loop. The source flux was adjusted with the current $i_{f}$ applied on the SLED source control. Our previous experiments were conducted in high flux regime for which $i_{f}$ is set to 25\,$\mu$A, serving as a reference in terms of source flux. In the following, we present our estimates of the integrated flux in the core of the non-coronagraphic image within a photometric aperture of 2.44\,$\lambda$/D diameter for a given $i_{f}$. The source flux can then be expressed with respect to the reference source flux in relative magnitude, $\Delta$mag.

\begin{figure}[!ht]
    \centering
    \includegraphics[width=\linewidth]{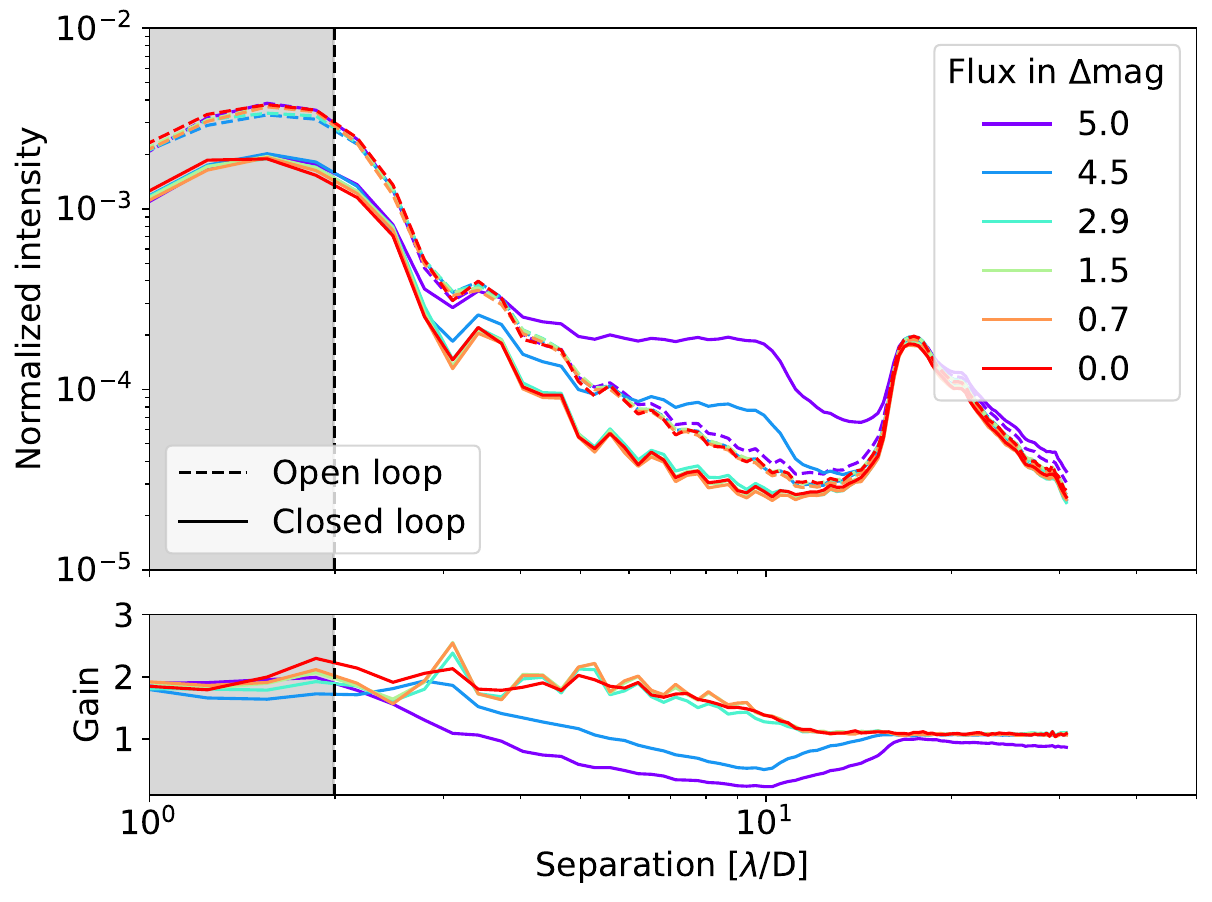}
    \caption{Contrast in the coronagraphic images with the ZWFS control loop for different source flux. \textbf{Top}: Normalized azimuthal averaged intensity profile of the coronagraphic images produced with the ZWFS wavefront control in open loop (dashed line) and closed loop (solid line) as a function of the angular separation for different source flux in relative magnitude $\Delta$mag. The grey area with dashed line delimits the projected FPM size. The AO residuals are based on the VLT/SPHERE characteristics with a 6-mag natural guide star and median observing conditions with 10\,m.s$^{-1}$ and 0.7" seeing. In closed loop, the ZWFS-based second-stage AO controls 350 KL modes and runs with an integrator using a loop gain of 0.8. \textbf{Bottom}: Contrast gain provided by the ZWFS-based wavefront control between the open and closed loop operations for different source flux.}
    \label{fig:cor_prf_sourceflux}
\end{figure}

Figure \ref{fig:cor_prf_sourceflux} shows the intensity profiles of the coronagraphic images in open and closed loops for different $\Delta$mag. The contrast performance is stable for $\Delta$mag up to 2.9. At larger relative magnitudes, the contrast degrades in the second-stage controlled region, showing a loss of efficiency for our scheme. In control loop operation, this effect relates to noise amplification at all the spatial frequencies, translating into contrast degradation in the area of the coronagraphic image within the ZWFS-based control radius. The result allows for an estimate of the source flux limit at which our second-stage AO scheme can operate efficiently. 

\begin{figure}
    \centering
    \includegraphics[width=\linewidth]{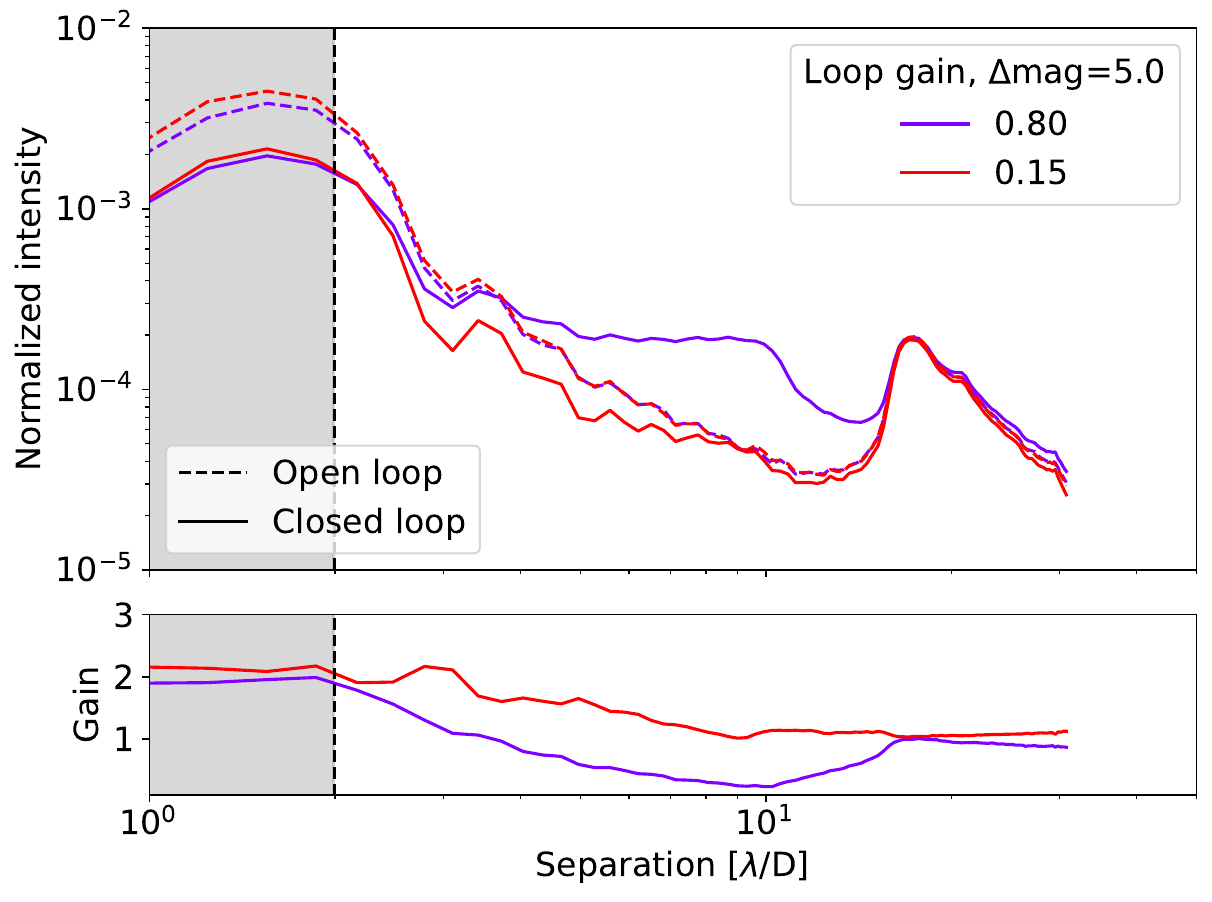}
    \caption{Contrast in the coronagraphic images with the ZWFS control loop and different gains for low source flux conditions. \textbf{Top}: Normalized azimuthal averaged intensity profile of the coronagraphic images produced with the ZWFS wavefront control in open loop (dashed line) and closed loop (solid line) as a function of the angular separation for a source flux with $\Delta$mag=5.0 and an integrator loop gain of 0.80 (purple) and 0.15 (red). The grey area with dashed line delimits the projected FPM size. The AO residuals are based on the VLT/SPHERE characteristics with a 6-mag natural guide star and median observing conditions with 10\,m.s$^{-1}$ and 0.7" seeing. In closed loop, the ZWFS-based second stage AO controls 350 KL modes. \textbf{Bottom}: Contrast gain provided by the ZWFS-based wavefront control between the open and closed loop operations for different loop gains.}
    \label{fig:cor_prf_sourceflux_00035}
\end{figure}

To address the noise amplification at the low flux regime, we performed tests on our control loop by adjusting the integrator gain. Such a procedure replicates the on-sky operation with a real AO system to ensure a stable closed loop in low flux regime. Figure \ref{fig:cor_prf_sourceflux_00035} shows an example for different loop gains with $\Delta$mag of 5.04. The noise amplification in the coronagraphic images for an integrator gain of 0.80 is removed by adjusting the loop gain to 0.15, leading a contrast gain that is comparable to the one obtained at the highest source flux in our experiments ($\Delta$mag=0.0). Our ZWFS control loop works in low flux regime and possible noise amplification is eliminated by adjusting the integrator gain.

\section{Comparison with the PWFS control loop}
The PWFS control loop is often considered as the baseline for the second-stage AO loop in exoplanet imagers. We compare our scheme with the PWFS control loop to determine some preliminary functioning points of interest for the ZWFS control loop for exoplanet observations with high-contrast instruments assisted with XAO systems. 

\subsection{Low flux regime}
The ZWFS is known to be twice more sensitive sensor than the PWFS in the unmodulated configuration \citep{Guyon2005} since its signal projects onto a single pupil, while PWFS splits the signal into four pupil images. This sensitivity gain is even larger when the comparison is made with the PWFS in modulated configuration. This advantage proves interesting in low flux regimes such as the observation of planetary companions around faint targets. To explore this possible benefit, we run an experiment with the same source flux and the same observing conditions for the ZWFS and the PWFS control loops. The parameters $\Delta$mag, $\beta$, and $v$ were set to 5.0, 0.7", and 10\,m.s$^{-1}$. For both configurations, we adjusted the loop gain to enable the control schemes to run without introducing noise amplification in the controlled region. The optimal integrator gain is found at 0.15 for the ZWFS and 0.03 for the PWFS in modulated configuration. 

\begin{figure}[!ht]
    \centering
    \includegraphics[width=\linewidth]{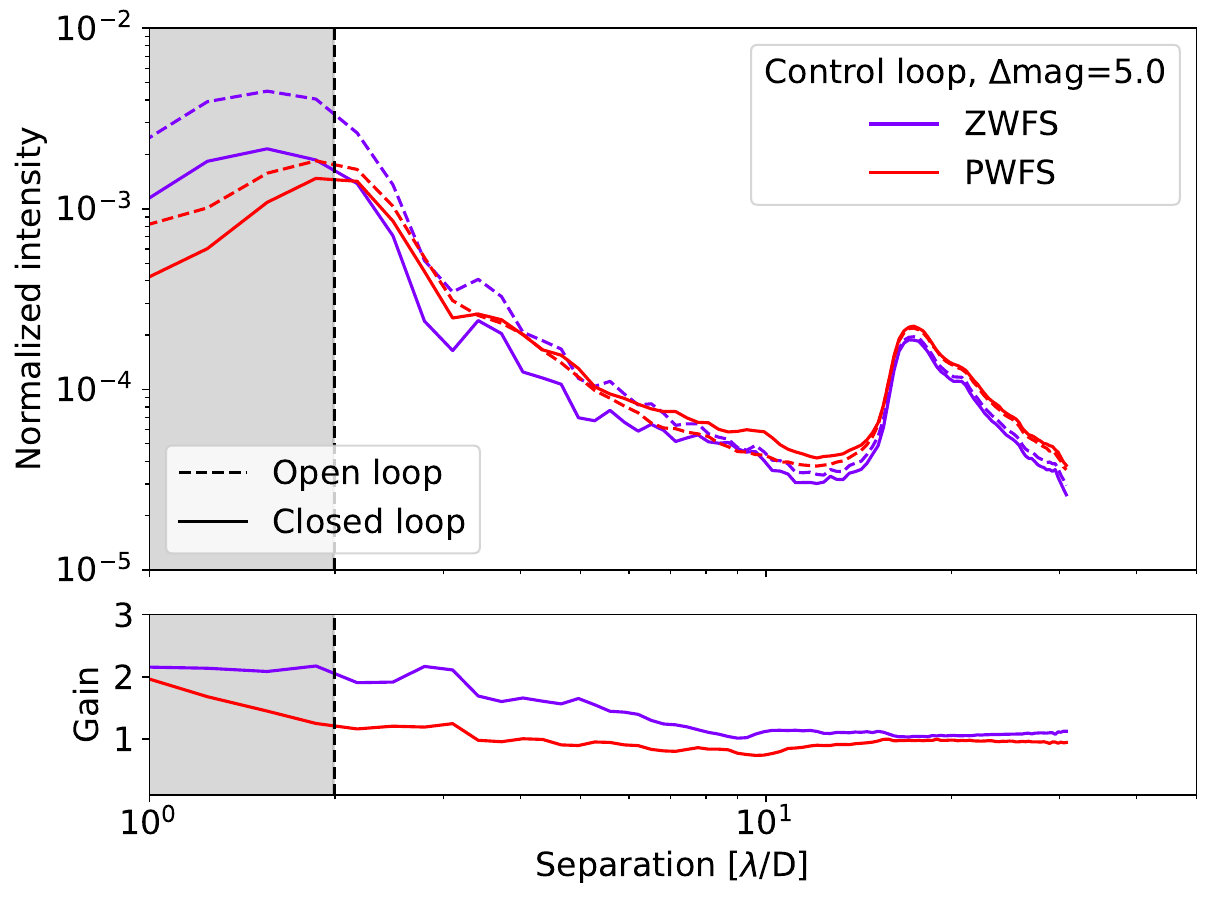}
    \caption{Contrast in the coronagraphic images with the ZWFS and PWFS control loops for low source flux conditions. \textbf{Top}: Normalized azimuthal averaged intensity profile of the coronagraphic images produced for the control loop with the ZWFS (purple) and the PWFS (red) in open loop (dashed line) and closed loop (solid line) as a function of the angular separation for a source flux with $\Delta$mag=$5.0$. The grey area with dashed line delimits the projected FPM size. The AO residuals are based on the VLT/SPHERE characteristics with a 6-mag natural guide star and median observing conditions with 10\,m.s$^{-1}$ and 0.7" seeing. In closed loop, the second stage AO controls 350 KL modes and runs with an integrator using an optimal gain. \textbf{Bottom}: Contrast gain provided by the different control loops between the open and closed loop operations.}
    \label{fig:cor_prf_sourceflux_zwfs-pwfs}
\end{figure}

Figure \ref{fig:cor_prf_sourceflux_zwfs-pwfs} represents the intensity profiles of the coronagraphic images in open and closed loop for both ZWFS and PWFS control schemes. The open loop profiles exhibit differences which are related to the different static aberrations that are applied on the DM in both cases. In the PWFS case, the flat response on the sensor is based on the by-eye optimization of the coefficients related to some first low-order Zernike modes. For the ZWFS configuration, we perform a two-step calibration as detailed in Sect. \ref{sec:protocol}. The wavefront pattern on the sensor results from a DM shape which produces a smooth and homogeneous intensity in the ZWFS pupil, offering a more robust control loop with this sensor, see Sect. \ref{subsec:calibration}. The flat intensity in the ZWFS pupil corresponds to wavefront errors that are non null, leading to some residual static low-order aberrations and speckles on the science camera. In an open loop, the PWFS control loop therefore offers a coronagraphic image with slightly deeper contrast than the ZWFS control loop. 

After closing the loop, the ZWFS control loop benefits from the sensitivity of its sensor with respect to the PWFS case and corrects for some of the atmospheric residuals, leading to contrast improvement in the coronagraphic image. In comparison, the PWFS control loop needs to use a very low gain to avoid noise amplification and hardly corrects for atmospheric residuals, leading to a limited contrast gain in the coronagraphic image. In summary, for the same flux and an optimal integrator gain, the ZWFS control loop offers slightly deeper contrasts and larger contrast gains than the modulated PWFS, confirming the benefit in sensitivity provided by the ZWFS over the PWFS. 

As this experiment presents its own limits, it is obviously risky to draw hasty conclusions. Thus, in this very preliminary test, the low flux configuration was set assuming the absence of impact on the correction of the first stage. In real life, the first-stage XAO system will see its performance limited in low flux regime and will most likely provide images with a moderate wavefront correction to the second stage AO. This situation will lead to aberration amplitudes that will certainly impact the ability of the second stage to correct for the AO residuals. In this scenario, the ZWFS control loop will be altered by the limited capture range of its sensor. Further studies therefore needs to be performed to compensate for these large residuals with the ZWFS control loop. Strategies such as the use of non-linear reconstructors \citep[e.g.,][]{Haffert2024} are promising options to run efficiently the ZWFS but this is out of the scope of the paper and will be addressed in future experiments.

Still, the control loop with ZWFS constitutes a complementary solution to the adopted second-stage baseline with the PWFS. Further investigations can reveal the benefits of working with a control loop based on a sensor with one of the highest sensitivities. 

\subsection{Control loop speed}
The PWFS sees its wavefront correction improved by modulating the source signal on the sensor. An appropriate set up is required to introduce the signal modulation on the pyramid. On the envisioned upgrade for VLT/SPHERE, the modulation runs with a tip/tilt platform at a speed up to 3\,kHz with an angular radius modulation of 3\,$\lambda/D$. Such a temporal frequency can constitute a limitation to run the second-stage control loop at larger frequencies. In contrast, the ZWFS control loop requires no modulation and therefore, our scheme can theoretically run at extremely high speed, possibly tens of kHz, assuming a sensor camera with a fast frame rate. 

To check this assumption, we run a dummy experiment in GHOST with $v$ of 34\,m.s$^{-1}$ and $\beta$ of 0.7". The SLM injects AO residuals at a frame rate of 350\,Hz and our control loop runs at frequencies of 350\,Hz and 5000\,Hz. Our tests lead to similar results in both cases, showing that the ZWFS can run at speed of nearly 15 times the standard frequency without any issue.

While our experiment is somehow unrealistic since the turbulence runs at 350Hz and the loop runs at 5000\,Hz, the turbulence does not evolve in-between loop corrections. However, the purpose here is to show that the ZWFS control loop can possibly run at extremely high speeds. While the modulated PWFS is limited by the modulation mirror to $\sim$3\,kHz, we show here that the ZWFS lifts this limitation and could be run on our off-the-shelf WFS and RTC hardware with 5\,kHz. Another feature of the ZWFS is that it only observes one pupil image, so the detector area is four times smaller than for the PWFS. Then, possible frame-rates are up to four times faster for a given camera depending on the readout concept.

\section{Discussion and conclusions}
We present a first proof of concept for the ZWFS-based second-stage AO loop, an alternative solution of the PWFS analog, to support a single-stage XAO system for coronagraphic observations of circumstellar environments. Our tests on GHOST validate the ability of our approach to reduce the AO residuals left by the first stage XAO loop, leading to contrast gains with a factor of at least 2 in the coronagraphic image of an observed point source. While encouraging, this gain of 2 is limited by the presence of quasi-static speckles and coronagraphic residuals in our experiment. The AO residuals are reduced by much larger factors, which would pay off in terms of contrast gains with a more efficient coronagraph and NCPA calibration scheme.

Our experiment shows that the ZWFS control loop provides similar contrast gains as the PWFS second-stage AO loop for the same observing conditions. As the ZWFS is twice more sensitive than the PWFS, the corresponding control loop scheme features a more robust contrast gain in low-flux regime, proving interesting for the observation of planetary companions around faint nearby stars. The PWFS can operate in modulated mode to improve the wavefront error correction of its control loop. This configuration presents a maximum speed \citep[e.g., 3\,kHz for the tip-tilt mirror modulation in the envisioned VLT/SPHERE+ upgrade][]{Boccaletti2022}, which limits the AO correction frequency of the PWFS-based control loop in modulated mode. The ZWFS control loop is free of modulation aspects, enabling wavefront control at speeds of up to 5 kHz or higher depending on the WFS camera and RTC hardware, provided that the natural guide star is bright enough for wavefront sensing. Such correction rates pave the way for wavefront correction in very high wind speed conditions and, therefore, a possible extension of the useful observing time and science return for the observation of exoplanetary systems. 

In our experiment, the ultimate contrast is mainly limited by the diffraction residuals from the classical Lyot coronagraph and by the NCPA existing between the control loop and the science path in the testbed. The first limitation can be removed by upgrading the starlight suppression system with a more elaborated coronagraph design to provide deeper contrast \citep[see, e.g., reviews by][]{Guyon2006,Ruane2018b}. The second limitation can be addressed by estimating the aberrations in science path, for instance, by using  some non-invasive strategies (e.g., phase diversity, differential optical transfer function) to calibrate the static errors, and then modifying the reference slopes in the calibration matrix. A preliminary estimate of the aberrations was already performed with the Fast \& Furious phase diversity algorithm in this paper. Additional real-time control strategies might be required to compensate for the quasi-static aberrations in GHOST.

The validation of our ZWFS-based approach is performed in monochromatic light. In real-life facilities, wide spectral band filters are often used to maximize the amount of photons on the WFS and improve the wavefront correction. Theoretical studies have already demonstrated the efficiency of the ZWFS for measuring aberrations in broadband light \citep{N'Diaye2013a,Haffert2024}. In the forthcoming tests, we plan to demonstrate the capability of ZWFS control loop to work in broadband light by installing a white light source and filters on GHOST. 

In AO systems, photon sharing between the wavefront sensing path and the science path often requires an adequate balance between both arms to ensure an optimal wavefront error correction without jeopardizing the observation of the faintest substellar mass companions. While the ZWFS is well-known for its high sensitivity, recent approaches have emerged to increase the sensor sensitivity even further by adjusting the mask size \citep{Chambouleyron2021} or by combining the sensor with an appropriate entrance pupil apodization \citep{Haffert2023}. These compelling approaches can push the wavefront correction further for the ZWFS control loop, making it an even more attractive solution for a two-stage AO loop. 

The ZWFS control loop relies on the linear capture range of the sensor to correct for the wavefront errors left from the first XAO stage. Since the ZWFS is renowned for its limited capture range, new flavors have recently emerged for this sensor to extend its capture range of the sensor further. These include  the vector ZWFS \citep{Doelman2019},  phase-shifting ZWFS \citep{Wallace2011},  combination of the ZWFS with phase diversity algorithms \citep{Haffert2024}, or the use of the ZWFS in multi-wavelength approach \citep{Vigan2011,Haffert2024}. Non-linear reconstructors based on neural network also constitute another promising avenue to extend the capture range of our approach, as it was recently proposed with the unmodulated PWFS \citep{Landman2024}. These schemes offer promising prospects to extend the sensor capture range further and broaden the use of the ZWFS wavefront control. 

In our study, the ZWFS control loop works in standalone mode, meaning the absence of interaction between the loops of the two AO stages. Command laws integrating the information of both loops represent a rising path to derive new approaches for the two-stage AO loop and maximize the overall correction of the wavefront errors due to the atmospheric turbulence. In this context, the temporal controller constitutes another key aspect to investigate further for the ZWFS control loop. Our current study relies on the use of an integrator in standalone mode. In future works, we will investigate more advanced command laws based on predictive control algorithms \citep{Poyneer2007,Males2018} or reinforcement learning methods \citep{Nousiainen2024} to further enhance the wavefront correction regime, along with the speed and accuracy of the ZWFS control loop in a standalone and in an integrated version with the first AO stage.

Our current study only considers atmospheric turbulence residuals at the output of the first AO stage. Other artifacts such as low-wind effects \citep{Sauvage2016,Milli2018}, pupil fragmentation, or cophasing errors in segmented aperture telescopes represent other limiting factors which can impact the efficiency of the wavefront correction provided by our scheme. We have already run some preliminary tests for our ZWFS control loop in the presence of AO residuals and petalling effects, showing promising results and exciting prospects for high-contrast observations of exoplanets \citep{N'Diaye2023}. We will investigate these aspects further in the near future. 

Finally, on-sky demonstration is a crucial step to increase the readiness level of our approach. Some on-sky studies with the Provence Adaptive-optics PYramid RUn System (PAPYRUS) testbed at Observatoire de Haute Provence in France are currently on going to demonstrate the feasibility of the two-stage AO with a vector ZWFS in the second loop \citep{Cisse2023}. The expected VLT/SPHERE+ project serves as a technology demonstrator to prepare for XAO facilities with ELTs for exoplanet imaging and spectroscopy \citep{Boccaletti2022}. The PWFS is selected in the baseline for the second-stage AO loop in the SPHERE upgrade. We are currently investigating the implementation of a ZWFS as an additional option to the PWFS to enlarge the wavefront correction options, while exploring suitable operating modes that could be left uncovered by the PWFS option and maximizing the science return for exoplanet imagers with ELTs. The exploration and in-lab validation of multiple and complementary second-stage AO strategies will be an asset for the most challenging observations of exoplanets from the ground, such as Earth-like planets around M dwarfs with ELTs \citep{Kasper2021,Fitzgerald2022,Kautz2023}. 

\begin{acknowledgements}
This work was supported by the Action Spécifique Haute Résolution Angulaire (ASHRA) of CNRS/INSU co-funded by CNES. MN acknowledges support from Observatoire de la Côte d'Azur and Laboratoire Lagrange through the 2022 BQR OCA and 2023 BQR Lagrange programs for the manufacturing of Zernike phase masks and the missions to ESO Garching. AV acknowledges funding from the European Research Council (ERC) under the European Union’s Horizon 2020 research and innovation programme, grant agreement No. 757561 (HiRISE).
\end{acknowledgements}

%-------------------------------------------------------------------
\bibliographystyle{aa} % style aa.bst
\bibliography{2024_mndiaye_biblio} % your references Yourfile.bib

\begin{thebibliography}{80}
\expandafter\ifx\csname natexlab\endcsname\relax\def\natexlab#1{#1}\fi

\bibitem[{{Ahn} {et~al.}(2021){Ahn}, {Guyon}, {Lozi}, {Vievard}, {Deo}, {Skaf}, {Belikov}, {Bos}, {Bottom}, {Currie}, {Frazin}, {V. Gorkom}, {Groff}, {Haffert}, {Jovanovic}, {Kawahara}, {Kotani}, {Males}, {Martinache}, {Mazin}, {Miller}, {Norris}, {Rodack}, \& {Wong}}]{Ahn2021}
{Ahn}, K., {Guyon}, O., {Lozi}, J., {et~al.} 2021, in Society of Photo-Optical Instrumentation Engineers (SPIE) Conference Series, Vol. 11823, Techniques and Instrumentation for Detection of Exoplanets X, ed. S.~B. {Shaklan} \& G.~J. {Ruane}, 1182303

\bibitem[{{Bailey} {et~al.}(2023){Bailey}, {Bendek}, {Monacelli}, {Baker}, {Bedrosian}, {Cady}, {Douglas}, {Groff}, {Hildebrandt}, {Kasdin}, {Krist}, {Macintosh}, {Mennesson}, {Morrissey}, {Poberezhskiy}, {Subedi}, {Rhodes}, {Roberge}, {Ygouf}, {Zellem}, {Zhao}, \& {Zimmerman}}]{Bailey2023}
{Bailey}, V.~P., {Bendek}, E., {Monacelli}, B., {et~al.} 2023, in Society of Photo-Optical Instrumentation Engineers (SPIE) Conference Series, Vol. 12680, Society of Photo-Optical Instrumentation Engineers (SPIE) Conference Series, 126800T

\bibitem[{{Beuzit} {et~al.}(2019){Beuzit}, {Vigan}, {Mouillet}, {Dohlen}, {Gratton}, {Boccaletti}, {Sauvage}, {Schmid}, {Langlois}, {Petit}, {Baruffolo}, {Feldt}, {Milli}, {Wahhaj}, {Abe}, {Anselmi}, {Antichi}, {Barette}, {Baudrand}, {Baudoz}, {Bazzon}, {Bernardi}, {Blanchard}, {Brast}, {Bruno}, {Buey}, {Carbillet}, {Carle}, {Cascone}, {Chapron}, {Charton}, {Chauvin}, {Claudi}, {Costille}, {De Caprio}, {de Boer}, {Delboulb{\'e}}, {Desidera}, {Dominik}, {Downing}, {Dupuis}, {Fabron}, {Fantinel}, {Farisato}, {Feautrier}, {Fedrigo}, {Fusco}, {Gigan}, {Ginski}, {Girard}, {Giro}, {Gisler}, {Gluck}, {Gry}, {Henning}, {Hubin}, {Hugot}, {Incorvaia}, {Jaquet}, {Kasper}, {Lagadec}, {Lagrange}, {Le Coroller}, {Le Mignant}, {Le Ruyet}, {Lessio}, {Lizon}, {Llored}, {Lundin}, {Madec}, {Magnard}, {Marteaud}, {Martinez}, {Maurel}, {M{\'e}nard}, {Mesa}, {M{\"o}ller-Nilsson}, {Moulin}, {Moutou}, {Orign{\'e}}, {Parisot}, {Pavlov}, {Perret}, {Pragt}, {Puget}, {Rabou}, {Ramos}, {Reess}, {Rigal}, {Rochat}, {Roelfsema}, {Rousset},
  {Roux}, {Saisse}, {Salasnich}, {Santambrogio}, {Scuderi}, {Segransan}, {Sevin}, {Siebenmorgen}, {Soenke}, {Stadler}, {Suarez}, {Tiph{\`e}ne}, {Turatto}, {Udry}, {Vakili}, {Waters}, {Weber}, {Wildi}, {Zins}, \& {Zurlo}}]{Beuzit2019}
{Beuzit}, J.~L., {Vigan}, A., {Mouillet}, D., {et~al.} 2019, \aap, 631, A155

\bibitem[{{Bloemhof} \& {Wallace}(2003)}]{Bloemhof2003}
{Bloemhof}, E.~E. \& {Wallace}, J.~K. 2003, in SPIE, Vol. 5169, 309--320

\bibitem[{{Boccaletti} {et~al.}(2020){Boccaletti}, {Chauvin}, {Mouillet}, {Absil}, {Allard}, {Antoniucci}, {Augereau}, {Barge}, {Baruffolo}, {Baudino}, {Baudoz}, {Beaulieu}, {Benisty}, {Beuzit}, {Bianco}, {Biller}, {Bonavita}, {Bonnefoy}, {Bos}, {Bouret}, {Brandner}, {Buchschache}, {Carry}, {Cantalloube}, {Cascone}, {Carlotti}, {Charnay}, {Chiavassa}, {Choquet}, {Clenet}, {Crida}, {De Boer}, {De Caprio}, {Desidera}, {Desert}, {Delisle}, {Delorme}, {Dohlen}, {Doelman}, {Dominik}, {Orazi}, {Dougados}, {Doute}, {Fedele}, {Feldt}, {Ferreira}, {Fontanive}, {Fusco}, {Galicher}, {Garufi}, {Gendron}, {Ghedina}, {Ginski}, {Gonzalez}, {Gratadour}, {Gratton}, {Guillot}, {Haffert}, {Hagelberg}, {Henning}, {Huby}, {Janson}, {Kamp}, {Keller}, {Kenworthy}, {Kervella}, {Kral}, {Kuhn}, {Lagadec}, {Laibe}, {Langlois}, {Lagrange}, {Launhardt}, {Leboulleux}, {Le Coroller}, {Li Causi}, {Loupias}, {Maire}, {Marleau}, {Martinache}, {Martinez}, {Mary}, {Mattioli}, {Mazoyer}, {Meheut}, {Menard}, {Mesa}, {Meunier}, {Miguel}, {Milli},
  {Min}, {Molliere}, {Mordasini}, {Moretto}, {Mugnier}, {Muro Arena}, {Nardetto}, {Diaye}, {Nesvadba}, {Pedichini}, {Pinilla}, {Por}, {Potier}, {Quanz}, {Rameau}, {Roelfsema}, {Rouan}, {Rigliaco}, {Salasnich}, {Samland}, {Sauvage}, {Schmid}, {Segransan}, {Snellen}, {Snik}, {Soulez}, {Stadler}, {Stam}, {Tallon}, {Thebault}, {Thiebaut}, {Tschudi}, {Udry}, {van Holstein}, {Vernazza}, {Vidal}, {Vigan}, {Waters}, {Wildi}, {Willson}, {Zanutta}, {Zavagno}, \& {Zurlo}}]{Boccaletti2020}
{Boccaletti}, A., {Chauvin}, G., {Mouillet}, D., {et~al.} 2020, arXiv e-prints, arXiv:2003.05714

\bibitem[{{Boccaletti} {et~al.}(2022){Boccaletti}, {Chauvin}, {Wildi}, {Milli}, {Stadler}, {Diolaiti}, {Gratton}, {Vidal}, {Loupias}, {Langlois}, {Cantalloube}, {N'Diaye}, {Gratadour}, {Ferreira}, {Tallon}, {Mazoyer}, {Segransan}, {Mouillet}, {Beuzit}, {Bonnefoy}, {Galicher}, {Vigan}, {Snellen}, {Feldt}, {Desidera}, {Rousseau}, {Baruffolo}, {Goulas}, {Baudoz}, {Bechet}, {Benisty}, {Bianco}, {Carry}, {Cascone}, {Charnay}, {Choquet}, {Christiaens}, {Cortecchia}, {Di Capprio}, {De Rosa}, {Desgrange}, {D'Orazi}, {Dout{\'e}}, {Frangiamore}, {Gendron}, {Ginski}, {Huby}, {Keller}, {Kulcs{\'a}r}, {Landman}, {Lagarde}, {Lagadec}, {Lagrange}, {Lombini}, {Kasper}, {M{\'e}nard}, {Magnard}, {Malaguti}, {Maurel}, {Mesa}, {Morgante}, {Pantin}, {Pichon}, {Potier}, {Rabou}, {Rochat}, {Terenzi}, {Thi{\'e}baut}, {Tallon-Bosc}, {Raynaud}, {Rouan}, {Sevin}, {Schiavone}, {Schrieber}, \& {Zanutta}}]{Boccaletti2022}
{Boccaletti}, A., {Chauvin}, G., {Wildi}, F., {et~al.} 2022, in Society of Photo-Optical Instrumentation Engineers (SPIE) Conference Series, Vol. 12184, Ground-based and Airborne Instrumentation for Astronomy IX, ed. C.~J. {Evans}, J.~J. {Bryant}, \& K.~{Motohara}, 121841S

\bibitem[{{Bos} {et~al.}(2020){Bos}, {Vievard}, {Wilby}, {Snik}, {Lozi}, {Guyon}, {Norris}, {Jovanovic}, {Martinache}, {Sauvage}, \& {Keller}}]{Bos2020}
{Bos}, S.~P., {Vievard}, S., {Wilby}, M.~J., {et~al.} 2020, \aap, 639, A52

\bibitem[{{Cantalloube} {et~al.}(2019){Cantalloube}, {Dohlen}, {Milli}, {Brandner}, \& {Vigan}}]{Cantalloube2019}
{Cantalloube}, F., {Dohlen}, K., {Milli}, J., {Brandner}, W., \& {Vigan}, A. 2019, The Messenger, 176, 25

\bibitem[{{Cantalloube} {et~al.}(2020){Cantalloube}, {Farley}, {Milli}, {Bharmal}, {Brandner}, {Correia}, {Dohlen}, {Henning}, {Osborn}, {Por}, {Su{\'a}rez Valles}, \& {Vigan}}]{Cantalloube2020}
{Cantalloube}, F., {Farley}, O.~J.~D., {Milli}, J., {et~al.} 2020, \aap, 638, A98

\bibitem[{{Cantalloube} {et~al.}(2018){Cantalloube}, {Por}, {Dohlen}, {Sauvage}, {Vigan}, {Kasper}, {Bharmal}, {Henning}, {Brandner}, {Milli}, {Correia}, \& {Fusco}}]{Cantalloube2018}
{Cantalloube}, F., {Por}, E.~H., {Dohlen}, K., {et~al.} 2018, \aap, 620, L10

\bibitem[{{Carter} {et~al.}(2021){Carter}, {Hinkley}, {Bonavita}, {Phillips}, {Girard}, {Perrin}, {Pueyo}, {Vigan}, {Gagn{\'e}}, \& {Skemer}}]{Carter2021}
{Carter}, A.~L., {Hinkley}, S., {Bonavita}, M., {et~al.} 2021, \mnras, 501, 1999

\bibitem[{{Cerpa-Urra} {et~al.}(2022){Cerpa-Urra}, {Kasper}, {Kulcs{\'a}r}, {Raynaud}, \& {Ta{\"\i}ssir Heritier}}]{Cerpa-Urra2022}
{Cerpa-Urra}, N., {Kasper}, M., {Kulcs{\'a}r}, C., {Raynaud}, H.-F., \& {Ta{\"\i}ssir Heritier}, C. 2022, Journal of Astronomical Telescopes, Instruments, and Systems, 8, 019001

\bibitem[{{Chambouleyron} {et~al.}(2021){Chambouleyron}, {Fauvarque}, {Sauvage}, {Dohlen}, {Levraud}, {Vigan}, {N'Diaye}, {Neichel}, \& {Fusco}}]{Chambouleyron2021}
{Chambouleyron}, V., {Fauvarque}, O., {Sauvage}, J.~F., {et~al.} 2021, \aap, 650, L8

\bibitem[{{Chauvin}(2024)}]{Chauvin2024}
{Chauvin}, G. 2024, Comptes Rendus Physique, 24, 139

\bibitem[{{Chilcote} {et~al.}(2022){Chilcote}, {Konopacky}, {Fitzsimmons}, {Hamper}, {Macintosh}, {Marois}, {Savransky}, {Soummer}, {V{\'e}ran}, {Agapito}, {Aleman}, {Ammons}, {Bonaglia}, {Boucher}, {Curliss}, {De Rosa}, {Do {\'O}}, {Dunn}, {Esposito}, {Filion}, {Kerley}, {Landry}, {Lardiere}, {Levinstein}, {Li}, {Limbach}, {Madurowicz}, {Maire}, {Millar-Blanchaer}, {Nickson}, {Nielsen}, {Nguyen}, {Nguyen}, {Peng}, {Perera}, {Perrin}, {Por}, {Poyneer}, {Pueyo}, {Rantakyr{\"o}}, {Sands}, {Spalding}, \& {Summey}}]{Chilcote2022}
{Chilcote}, J., {Konopacky}, Q., {Fitzsimmons}, J., {et~al.} 2022, in Society of Photo-Optical Instrumentation Engineers (SPIE) Conference Series, Vol. 12184, Ground-based and Airborne Instrumentation for Astronomy IX, ed. C.~J. {Evans}, J.~J. {Bryant}, \& K.~{Motohara}, 121841T

\bibitem[{Cisse {et~al.}(2023)Cisse, Muslimov, Taissir~Heritier, Chambouleyron, Fetick, Levraud, Sauvage, Neichel, \& Fusco}]{Cisse2023}
Cisse, M., Muslimov, E., Taissir~Heritier, C., {et~al.} 2023, in {Adaptive Optics for Extremely Large Telescopes 7th Edition}, {ONERA}, Avignon, France

\bibitem[{{Currie} {et~al.}(2023){Currie}, {Biller}, {Lagrange}, {Marois}, {Guyon}, {Nielsen}, {Bonnefoy}, \& {De Rosa}}]{Currie2023}
{Currie}, T., {Biller}, B., {Lagrange}, A., {et~al.} 2023, in Astronomical Society of the Pacific Conference Series, Vol. 534, Protostars and Planets VII, ed. S.~{Inutsuka}, Y.~{Aikawa}, T.~{Muto}, K.~{Tomida}, \& M.~{Tamura}, 799

\bibitem[{{Debes} {et~al.}(2019){Debes}, {Ren}, \& {Schneider}}]{Debes2019}
{Debes}, J.~H., {Ren}, B., \& {Schneider}, G. 2019, Journal of Astronomical Telescopes, Instruments, and Systems, 5, 035003

\bibitem[{{Doelman} {et~al.}(2019){Doelman}, {Fagginger Auer}, {Escuti}, \& {Snik}}]{Doelman2019}
{Doelman}, D.~S., {Fagginger Auer}, F., {Escuti}, M.~J., \& {Snik}, F. 2019, Optics Letters, 44, 17

\bibitem[{{Dohlen}(2004)}]{Dohlen2004}
{Dohlen}, K. 2004, in EAS Publications Series, ed. C.~{Aime} \& R.~{Soummer}, Vol.~12, 33--44

\bibitem[{{Dohlen} {et~al.}(2006){Dohlen}, {Langlois}, {Lanzoni}, {Mazzanti}, {Vigan}, {Montoya}, {Hernandez}, {Reyes}, {Surdej}, \& {Yaitskova}}]{Dohlen2006}
{Dohlen}, K., {Langlois}, M., {Lanzoni}, P., {et~al.} 2006, in SPIE, Vol. 6267

\bibitem[{Ferreira {et~al.}(2022)Ferreira, Bernard, Sevin, Doucet, \& Gratadour}]{Ferreira2022}
Ferreira, F., Bernard, J., Sevin, A., Doucet, N., \& Gratadour, D. 2022, in 2022 IEEE Workshop on Signal Processing Systems (SiPS), 1--6

\bibitem[{{Fitzgerald} {et~al.}(2022){Fitzgerald}, {Sallum}, {Millar-Blanchaer}, {Jensen-Clem}, {Hinz}, {Guyon}, {Wang}, {Mazin}, {Skemer}, {Chun}, {Males}, {Marois}, {Singh}, \& {Max}}]{Fitzgerald2022}
{Fitzgerald}, M.~P., {Sallum}, S., {Millar-Blanchaer}, M.~A., {et~al.} 2022, in Society of Photo-Optical Instrumentation Engineers (SPIE) Conference Series, Vol. 12184, Ground-based and Airborne Instrumentation for Astronomy IX, ed. C.~J. {Evans}, J.~J. {Bryant}, \& K.~{Motohara}, 1218426

\bibitem[{{Fusco} {et~al.}(2006){Fusco}, {Rousset}, {Sauvage}, {Petit}, {Beuzit}, {Dohlen}, {Mouillet}, {Charton}, {Nicolle}, {Kasper}, {Baudoz}, \& {Puget}}]{Fusco2006}
{Fusco}, T., {Rousset}, G., {Sauvage}, J.-F., {et~al.} 2006, Optics Express, 14, 7515

\bibitem[{{Fusco} {et~al.}(2016){Fusco}, {Sauvage}, {Mouillet}, {Costille}, {Petit}, {Beuzit}, {Dohlen}, {Milli}, {Girard}, {Kasper}, {Vigan}, {Suarez}, {Soenke}, {Downing}, {N'Diaye}, {Baudoz}, {Sevin}, {Baruffolo}, {Schmid}, {Salasnich}, {Hugot}, \& {Hubin}}]{Fusco2016}
{Fusco}, T., {Sauvage}, J.~F., {Mouillet}, D., {et~al.} 2016, in Society of Photo-Optical Instrumentation Engineers (SPIE) Conference Series, Vol. 9909, Adaptive Optics Systems V, ed. E.~{Marchetti}, L.~M. {Close}, \& J.-P. {V{\'e}ran}, 99090U

\bibitem[{{Galicher} \& {Mazoyer}(2024)}]{Galicher2024}
{Galicher}, R. \& {Mazoyer}, J. 2024, Comptes Rendus Physique, 24, 133

\bibitem[{{Gaudi} {et~al.}(2020){Gaudi}, {Seager}, {Mennesson}, {Kiessling}, {Warfield}, {Cahoy}, {Clarke}, {Domagal-Goldman}, {Feinberg}, {Guyon}, {Kasdin}, {Mawet}, {Plavchan}, {Robinson}, {Rogers}, {Scowen}, {Somerville}, {Stapelfeldt}, {Stark}, {Stern}, {Turnbull}, {Amini}, {Kuan}, {Martin}, {Morgan}, {Redding}, {Stahl}, {Webb}, {Alvarez-Salazar}, {Arnold}, {Arya}, {Balasubramanian}, {Baysinger}, {Bell}, {Below}, {Benson}, {Blais}, {Booth}, {Bourgeois}, {Bradford}, {Brewer}, {Brooks}, {Cady}, {Caldwell}, {Calvet}, {Carr}, {Chan}, {Cormarkovic}, {Coste}, {Cox}, {Danner}, {Davis}, {Dewell}, {Dorsett}, {Dunn}, {East}, {Effinger}, {Eng}, {Freebury}, {Garcia}, {Gaskin}, {Greene}, {Hennessy}, {Hilgemann}, {Hood}, {Holota}, {Howe}, {Huang}, {Hull}, {Hunt}, {Hurd}, {Johnson}, {Kissil}, {Knight}, {Kolenz}, {Kraus}, {Krist}, {Li}, {Lisman}, {Mandic}, {Mann}, {Marchen}, {Marrese-Reading}, {McCready}, {McGown}, {Missun}, {Miyaguchi}, {Moore}, {Nemati}, {Nikzad}, {Nissen}, {Novicki}, {Perrine}, {Pineda}, {Polanco},
  {Putnam}, {Qureshi}, {Richards}, {Eldorado Riggs}, {Rodgers}, {Rud}, {Saini}, {Scalisi}, {Scharf}, {Schulz}, {Serabyn}, {Sigrist}, {Sikkia}, {Singleton}, {Shaklan}, {Smith}, {Southerd}, {Stahl}, {Steeves}, {Sturges}, {Sullivan}, {Tang}, {Taras}, {Tesch}, {Therrell}, {Tseng}, {Valente}, {Van Buren}, {Villalvazo}, {Warwick}, {Webb}, {Westerhoff}, {Wofford}, {Wu}, {Woo}, {Wood}, {Ziemer}, {Arney}, {Anderson}, {Ma{\'\i}z-Apell{\'a}niz}, {Bartlett}, {Belikov}, {Bendek}, {Cenko}, {Douglas}, {Dulz}, {Evans}, {Faramaz}, {Feng}, {Ferguson}, {Follette}, {Ford}, {Garc{\'\i}a}, {Geha}, {Gelino}, {G{\"o}tberg}, {Hildebrandt}, {Hu}, {Jahnke}, {Kennedy}, {Kreidberg}, {Isella}, {Lopez}, {Marchis}, {Macri}, {Marley}, {Matzko}, {Mazoyer}, {McCandliss}, {Meshkat}, {Mordasini}, {Morris}, {Nielsen}, {Newman}, {Petigura}, {Postman}, {Reines}, {Roberge}, {Roederer}, {Ruane}, {Schwieterman}, {Sirbu}, {Spalding}, {Teplitz}, {Tumlinson}, {Turner}, {Werk}, {Wofford}, {Wyatt}, {Young}, \& {Zellem}}]{HabExreport}
{Gaudi}, B.~S., {Seager}, S., {Mennesson}, B., {et~al.} 2020, arXiv e-prints, arXiv:2001.06683

\bibitem[{{Guyon}(2005)}]{Guyon2005}
{Guyon}, O. 2005, \apj, 629, 592

\bibitem[{{Guyon}(2018)}]{Guyon2018}
{Guyon}, O. 2018, \araa, 56, 315

\bibitem[{{Guyon} {et~al.}(2006){Guyon}, {Pluzhnik}, {Kuchner}, {Collins}, \& {Ridgway}}]{Guyon2006}
{Guyon}, O., {Pluzhnik}, E.~A., {Kuchner}, M.~J., {Collins}, B., \& {Ridgway}, S.~T. 2006, \apjs, 167, 81

\bibitem[{{Haffert} {et~al.}(2023){Haffert}, {Males}, \& {Guyon}}]{Haffert2023}
{Haffert}, S., {Males}, J., \& {Guyon}, O. 2023, in Adaptive Optics for Extremely Large Telescopes (AO4ELT7), 79

\bibitem[{{Haffert}(2024)}]{Haffert2024}
{Haffert}, S.~Y. 2024, \aap, 683, A113

\bibitem[{Heritier {et~al.}(2023)Heritier, Verinaud, \& Correia}]{Heritier2023}
Heritier, C.~T., Verinaud, C., \& Correia, C.~M. 2023, in {Adaptive Optics for Extremely Large Telescopes 7th Edition}, {ONERA}, Avignon, France

\bibitem[{{Hinkley} {et~al.}(2022){Hinkley}, {Carter}, {Ray}, {Skemer}, {Biller}, {Choquet}, {Millar-Blanchaer}, {Sallum}, {Miles}, {Whiteford}, {Patapis}, {Perrin}, {Pueyo}, {Schneider}, {Stapelfeldt}, {Wang}, {Ward-Duong}, {Bowler}, {Boccaletti}, {Girard}, {Hines}, {Kalas}, {Kammerer}, {Kervella}, {Leisenring}, {Pantin}, {Zhou}, {Meyer}, {Liu}, {Bonnefoy}, {Currie}, {McElwain}, {Metchev}, {Wyatt}, {Absil}, {Adams}, {Barman}, {Baraffe}, {Bonavita}, {Booth}, {Bryan}, {Chauvin}, {Chen}, {Danielski}, {De Furio}, {Factor}, {Fitzgerald}, {Fortney}, {Grady}, {Greenbaum}, {Henning}, {Hoch}, {Janson}, {Kennedy}, {Kenworthy}, {Kraus}, {Kuzuhara}, {Lagage}, {Lagrange}, {Launhardt}, {Lazzoni}, {Lloyd}, {Marino}, {Marley}, {Martinez}, {Marois}, {Matthews}, {Matthews}, {Mawet}, {Mazoyer}, {Phillips}, {Petrus}, {Quanz}, {Quirrenbach}, {Rameau}, {Rebollido}, {Rickman}, {Samland}, {Sargent}, {Schlieder}, {Sivaramakrishnan}, {Stone}, {Tamura}, {Tremblin}, {Uyama}, {Vasist}, {Vigan}, {Wagner}, \& {Ygouf}}]{Hinkley2022}
{Hinkley}, S., {Carter}, A.~L., {Ray}, S., {et~al.} 2022, \pasp, 134, 095003

\bibitem[{{Janin-Potiron} {et~al.}(2017){Janin-Potiron}, {N'Diaye}, {Martinez}, {Vigan}, {Dohlen}, \& {Carbillet}}]{Janin-Potiron2017}
{Janin-Potiron}, P., {N'Diaye}, M., {Martinez}, P., {et~al.} 2017, \aap, 603, A23

\bibitem[{Jensen-Clem {et~al.}(2012)Jensen-Clem, Wallace, \& Serabyn}]{Jensen-Clem2012}
Jensen-Clem, R., Wallace, J.~K., \& Serabyn, E. 2012, in 2012 IEEE Aerospace Conference, 1--7

\bibitem[{{Jovanovic} {et~al.}(2019){Jovanovic}, {Delorme}, {Bond}, {Cetre}, {Mawet}, {Echeverri}, {Wallace}, {Bartos}, {Lilley}, {Ragland}, {Ruane}, {Wizinowich}, {Chun}, {Wang}, {Wang}, {Fitzgerald}, {Matthews}, {Pezzato}, {Calvin}, {Millar-Blanchaer}, {Martin}, {Wetherell}, {Wang}, {Jacobson}, {Warmbier}, {Lockhart}, {Hall}, {Jensen-Clem}, \& {McEwen}}]{Jovanovic2019}
{Jovanovic}, N., {Delorme}, J.~R., {Bond}, C.~Z., {et~al.} 2019, in Techniques and Instrumentation for Detection of Exoplanets IX, ed. S.~B. Shaklan, Vol. 11117, International Society for Optics and Photonics (SPIE), 111170T

\bibitem[{{Jovanovic} {et~al.}(2015){Jovanovic}, {Martinache}, {Guyon}, {Clergeon}, {Singh}, {Kudo}, {Garrel}, {Newman}, {Doughty}, {Lozi}, {Males}, {Minowa}, {Hayano}, {Takato}, {Morino}, {Kuhn}, {Serabyn}, {Norris}, {Tuthill}, {Schworer}, {Stewart}, {Close}, {Huby}, {Perrin}, {Lacour}, {Gauchet}, {Vievard}, {Murakami}, {Oshiyama}, {Baba}, {Matsuo}, {Nishikawa}, {Tamura}, {Lai}, {Marchis}, {Duchene}, {Kotani}, \& {Woillez}}]{Jovanovic2015}
{Jovanovic}, N., {Martinache}, F., {Guyon}, O., {et~al.} 2015, \pasp, 127, 890

\bibitem[{{Kasper} {et~al.}(2021){Kasper}, {Cerpa Urra}, {Pathak}, {Bonse}, {Nousiainen}, {Engler}, {Heritier}, {Kammerer}, {Leveratto}, {Rajani}, {Bristow}, {Le Louarn}, {Madec}, {Str{\"o}bele}, {Verinaud}, {Glauser}, {Quanz}, {Helin}, {Keller}, {Snik}, {Boccaletti}, {Chauvin}, {Mouillet}, {Kulcs{\'a}r}, \& {Raynaud}}]{Kasper2021}
{Kasper}, M., {Cerpa Urra}, N., {Pathak}, P., {et~al.} 2021, The Messenger, 182, 38

\bibitem[{{Kasper} {et~al.}(2004){Kasper}, {Fedrigo}, {Looze}, {Bonnet}, {Ivanescu}, \& {Oberti}}]{Kasper2004}
{Kasper}, M., {Fedrigo}, E., {Looze}, D.~P., {et~al.} 2004, Journal of the Optical Society of America A, 21, 1004

\bibitem[{{Kautz} {et~al.}(2023){Kautz}, {Males}, {Close}, {Haffert}, {Guyon}, {Hedglen}, {Gasho}, {Durney}, {Noenickx}, {Fletcher}, {Coronado}, {Ford}, {Connors}, {Sullivan}, {Salanski}, {Kelly}, {Demers}, {Bouchez}, {Sitarski}, \& {Schurter}}]{Kautz2023}
{Kautz}, M., {Males}, J., {Close}, L., {et~al.} 2023, in Adaptive Optics for Extremely Large Telescopes (AO4ELT7), 61

\bibitem[{{Keller} {et~al.}(2012){Keller}, {Korkiakoski}, {Doelman}, {Fraanje}, {Andrei}, \& {Verhaegen}}]{Keller2012}
{Keller}, C.~U., {Korkiakoski}, V., {Doelman}, N., {et~al.} 2012, in Society of Photo-Optical Instrumentation Engineers (SPIE) Conference Series, Vol. 8447, Adaptive Optics Systems III, ed. B.~L. {Ellerbroek}, E.~{Marchetti}, \& J.-P. {V{\'e}ran}, 844721

\bibitem[{{Korkiakoski} {et~al.}(2014){Korkiakoski}, {Keller}, {Doelman}, {Kenworthy}, {Otten}, \& {Verhaegen}}]{Korkiakoski2014}
{Korkiakoski}, V., {Keller}, C.~U., {Doelman}, N., {et~al.} 2014, \ao, 53, 4565

\bibitem[{{Landman} {et~al.}(2024){Landman}, {Haffert}, {Males}, {Close}, {Foster}, {Van Gorkom}, {Guyon}, {Hedglen}, {Kautz}, {Kueny}, {Long}, {Lumbres}, {McEwen}, {McLeod}, \& {Schatz}}]{Landman2024}
{Landman}, R., {Haffert}, S.~Y., {Males}, J.~R., {et~al.} 2024, \aap, 684, A114

\bibitem[{{Lyot}(1932)}]{Lyot1932}
{Lyot}, B. 1932, Zeitschrift fur Astrophysik, 5, 73

\bibitem[{{Macintosh} {et~al.}(2014){Macintosh}, {Graham}, {Ingraham}, {Konopacky}, {Marois}, {Perrin}, {Poyneer}, {Bauman}, {Barman}, {Burrows}, {Cardwell}, {Chilcote}, {De Rosa}, {Dillon}, {Doyon}, {Dunn}, {Erikson}, {Fitzgerald}, {Gavel}, {Goodsell}, {Hartung}, {Hibon}, {Kalas}, {Larkin}, {Maire}, {Marchis}, {Marley}, {McBride}, {Millar-Blanchaer}, {Morzinski}, {Norton}, {Oppenheimer}, {Palmer}, {Patience}, {Pueyo}, {Rantakyro}, {Sadakuni}, {Saddlemyer}, {Savransky}, {Serio}, {Soummer}, {Sivaramakrishnan}, {Song}, {Thomas}, {Wallace}, {Wiktorowicz}, \& {Wolff}}]{Macintosh2014}
{Macintosh}, B., {Graham}, J.~R., {Ingraham}, P., {et~al.} 2014, Proceedings of the National Academy of Science, 111, 12661

\bibitem[{{Males} {et~al.}(2022){Males}, {Close}, {Haffert}, {Long}, {Hedglen}, {Pearce}, {Weinberger}, {Guyon}, {Knight}, {McLeod}, {Kautz}, {Van Gorkom}, {Lumbres}, {Schatz}, {Rodack}, {Gasho}, {Kueny}, \& {Foster}}]{Males2022}
{Males}, J.~R., {Close}, L.~M., {Haffert}, S., {et~al.} 2022, in Society of Photo-Optical Instrumentation Engineers (SPIE) Conference Series, Vol. 12185, Adaptive Optics Systems VIII, ed. L.~{Schreiber}, D.~{Schmidt}, \& E.~{Vernet}, 1218509

\bibitem[{{Males} \& {Guyon}(2018)}]{Males2018}
{Males}, J.~R. \& {Guyon}, O. 2018, Journal of Astronomical Telescopes, Instruments, and Systems, 4, 019001

\bibitem[{{Mawet} {et~al.}(2016){Mawet}, {Wizinowich}, {Dekany}, {Chun}, {Hall}, {Cetre}, {Guyon}, {Wallace}, {Bowler}, {Liu}, {Ruane}, {Serabyn}, {Bartos}, {Wang}, {Vasisht}, {Fitzgerald}, {Skemer}, {Ireland}, {Fucik}, {Fortney}, {Crossfield}, {Hu}, \& {Benneke}}]{Mawet2016}
{Mawet}, D., {Wizinowich}, P., {Dekany}, R., {et~al.} 2016, in Society of Photo-Optical Instrumentation Engineers (SPIE) Conference Series, Vol. 9909, Adaptive Optics Systems V, ed. E.~{Marchetti}, L.~M. {Close}, \& J.-P. {V{\'e}ran}, 99090D

\bibitem[{{Milli} {et~al.}(2018){Milli}, {Kasper}, {Bourget}, {Pannetier}, {Mouillet}, {Sauvage}, {Reyes}, {Fusco}, {Cantalloube}, {Tristam}, {Wahhaj}, {Beuzit}, {Girard}, {Mawet}, {Telle}, {Vigan}, \& {N'Diaye}}]{Milli2018}
{Milli}, J., {Kasper}, M., {Bourget}, P., {et~al.} 2018, in Society of Photo-Optical Instrumentation Engineers (SPIE) Conference Series, Vol. 10703, Adaptive Optics Systems VI, ed. L.~M. {Close}, L.~{Schreiber}, \& D.~{Schmidt}, 107032A

\bibitem[{{N'Diaye} {et~al.}(2010){N'Diaye}, {Dohlen}, {Cuevas}, {Lanzoni}, {Chemla}, {Chaumont}, {Soummer}, \& {Griffiths}}]{N'Diaye2010}
{N'Diaye}, M., {Dohlen}, K., {Cuevas}, S., {et~al.} 2010, \aap, 509, A8

\bibitem[{{N'Diaye} {et~al.}(2013){N'Diaye}, {Dohlen}, {Fusco}, \& {Paul}}]{N'Diaye2013a}
{N'Diaye}, M., {Dohlen}, K., {Fusco}, T., \& {Paul}, B. 2013, \aap, 555, A94

\bibitem[{{N'Diaye} {et~al.}(2016){N'Diaye}, {Soummer}, {Pueyo}, {Carlotti}, {Stark}, \& {Perrin}}]{N'Diaye2016}
{N'Diaye}, M., {Soummer}, R., {Pueyo}, L., {et~al.} 2016, \apj, 818, 163

\bibitem[{N'Diaye {et~al.}(2023)N'Diaye, Vigan, Engler, Kasper, Leveratto, Floriot, Marcos, Bailet, \& Dohlen}]{N'Diaye2023}
N'Diaye, M., Vigan, A., Engler, B., {et~al.} 2023, in {Adaptive Optics for Extremely Large Telescopes 7th Edition}, {ONERA}, Avignon, France

\bibitem[{{Nousiainen} {et~al.}(2024){Nousiainen}, {Engler}, {Kasper}, {Rajani}, {Helin}, {Heritier}, {Quanz}, \& {Glauser}}]{Nousiainen2024}
{Nousiainen}, J., {Engler}, B., {Kasper}, M., {et~al.} 2024, Journal of Astronomical Telescopes, Instruments, and Systems, 10, 019001

\bibitem[{{Oppenheimer} \& {Hinkley}(2009)}]{Oppenheimer2009}
{Oppenheimer}, B.~R. \& {Hinkley}, S. 2009, \araa, 47, 253

\bibitem[{{Perrin} {et~al.}(2003){Perrin}, {Sivaramakrishnan}, {Makidon}, {Oppenheimer}, \& {Graham}}]{Perrin2003}
{Perrin}, M.~D., {Sivaramakrishnan}, A., {Makidon}, R.~B., {Oppenheimer}, B.~R., \& {Graham}, J.~R. 2003, \apj, 596, 702

\bibitem[{{Pourcelot} {et~al.}(2022){Pourcelot}, {N'Diaye}, {Por}, {Laginja}, {Carbillet}, {Benard}, {Brady}, {Canas}, {Dohlen}, {Fowler}, {Lai}, {Maclay}, {McChesney}, {Noss}, {Perrin}, {Petrone}, {Pueyo}, {Redmond}, {Sahoo}, {Vigan}, {Will}, \& {Soummer}}]{Pourcelot2022}
{Pourcelot}, R., {N'Diaye}, M., {Por}, E.~H., {et~al.} 2022, \aap, 663, A49

\bibitem[{{Pourcelot} {et~al.}(2023){Pourcelot}, {Por}, {N'Diaye}, {Benard}, {Brady}, {Canas}, {Carbillet}, {Dohlen}, {Laginja}, {Lugten}, {Noss}, {Perrin}, {Petrone}, {Pueyo}, {Redmond}, {Sahoo}, {Vigan}, {Will}, \& {Soummer}}]{Pourcelot2023}
{Pourcelot}, R., {Por}, E.~H., {N'Diaye}, M., {et~al.} 2023, \aap, 672, A73

\bibitem[{{Pourcelot} {et~al.}(2021){Pourcelot}, {Vigan}, {Dohlen}, {Rouz{\'e}}, {Sauvage}, {El Morsy}, {Lopez}, {N'Diaye}, {Caillat}, {Choquet}, {Otten}, {Abbinanti}, {Balard}, {Carbillet}, {Blanchard}, {Hulin}, \& {Robert}}]{Pourcelot2021}
{Pourcelot}, R., {Vigan}, A., {Dohlen}, K., {et~al.} 2021, \aap, 649, A170

\bibitem[{{Poyneer} \& {Macintosh}(2004)}]{Poyneer2004}
{Poyneer}, L.~A. \& {Macintosh}, B. 2004, Journal of the Optical Society of America A, 21, 810

\bibitem[{{Poyneer} {et~al.}(2007){Poyneer}, {Macintosh}, \& {V{\'e}ran}}]{Poyneer2007}
{Poyneer}, L.~A., {Macintosh}, B.~A., \& {V{\'e}ran}, J.-P. 2007, Journal of the Optical Society of America A, 24, 2645

\bibitem[{{Ragazzoni}(1996)}]{Ragazzoni1996}
{Ragazzoni}, R. 1996, Journal of Modern Optics, 43, 289

\bibitem[{{Ruane} {et~al.}(2018){Ruane}, {Riggs}, {Mazoyer}, {Por}, {N'Diaye}, {Huby}, {Baudoz}, {Galicher}, {Douglas}, {Knight}, {Carlomagno}, {Fogarty}, {Pueyo}, {Zimmerman}, {Absil}, {Beaulieu}, {Cady}, {Carlotti}, {Doelman}, {Guyon}, {Haffert}, {Jewell}, {Jovanovic}, {Keller}, {Kenworthy}, {Kuhn}, {Miller}, {Sirbu}, {Snik}, {Wallace}, {Wilby}, \& {Ygouf}}]{Ruane2018b}
{Ruane}, G., {Riggs}, A., {Mazoyer}, J., {et~al.} 2018, in Society of Photo-Optical Instrumentation Engineers (SPIE) Conference Series, Vol. 10698, Space Telescopes and Instrumentation 2018: Optical, Infrared, and Millimeter Wave, ed. M.~{Lystrup}, H.~A. {MacEwen}, G.~G. {Fazio}, N.~{Batalha}, N.~{Siegler}, \& E.~C. {Tong}, 106982S

\bibitem[{{Ruane} {et~al.}(2020){Ruane}, {Wallace}, {Steeves}, {Prada}, {Seo}, {Bendek}, {Coker}, {Chen}, {Crill}, {Jewell}, {Kern}, {Marx}, {Poon}, {Redding}, {Riggs}, {Siegler}, \& {Zimmer}}]{Ruane2020}
{Ruane}, G., {Wallace}, J.~K., {Steeves}, J., {et~al.} 2020, Journal of Astronomical Telescopes, Instruments, and Systems, 6, 045005

\bibitem[{{Salama} {et~al.}(2024){Salama}, {Guthery}, {Chambouleyron}, {Jensen-Clem}, {Wallace}, {Delorme}, {Troy}, {Wenger}, {Echeverri}, {Finnerty}, {Jovanovic}, {Liberman}, {L{\'o}pez}, {Mawet}, {Morris}, {van Kooten}, {Wang}, {Wizinowich}, {Xin}, \& {Xuan}}]{Salama2024}
{Salama}, M., {Guthery}, C., {Chambouleyron}, V., {et~al.} 2024, \apj, 967, 171

\bibitem[{{Sauvage} {et~al.}(2016){Sauvage}, {Fusco}, {Lamb}, {Girard}, {Brinkmann}, {Guesalaga}, {Wizinowich}, {O'Neal}, {N'Diaye}, {Vigan}, {Mouillet}, {Beuzit}, {Kasper}, {Le Louarn}, {Milli}, {Dohlen}, {Neichel}, {Bourget}, {Haguenauer}, \& {Mawet}}]{Sauvage2016}
{Sauvage}, J.-F., {Fusco}, T., {Lamb}, M., {et~al.} 2016, in \procspie, Vol. 9909, Society of Photo-Optical Instrumentation Engineers (SPIE) Conference Series, 990916

\bibitem[{{Shi} {et~al.}(2016){Shi}, {Balasubramanian}, {Hein}, {Lam}, {Moore}, {Moore}, {Patterson}, {Poberezhskiy}, {Shields}, {Sidick}, {Tang}, {Truong}, {Wallace}, {Wang}, \& {Wilson}}]{Shi2016}
{Shi}, F., {Balasubramanian}, K., {Hein}, R., {et~al.} 2016, Journal of Astronomical Telescopes, Instruments, and Systems, 2, 011021

\bibitem[{{Steeves} {et~al.}(2020){Steeves}, {Wallace}, {Kettenbeil}, \& {Jewell}}]{Steeves2020}
{Steeves}, J., {Wallace}, J.~K., {Kettenbeil}, C., \& {Jewell}, J. 2020, Optica, 7, 1267

\bibitem[{{Surdej} {et~al.}(2010){Surdej}, {Yaitskova}, \& {Gonte}}]{Surdej2010}
{Surdej}, I., {Yaitskova}, N., \& {Gonte}, F. 2010, \ao, 49, 4052

\bibitem[{{The LUVOIR Team}(2019)}]{LUVOIRreport}
{The LUVOIR Team}. 2019, arXiv e-prints, arXiv:1912.06219

\bibitem[{{Traub} \& {Oppenheimer}(2010)}]{Traub2010}
{Traub}, W.~A. \& {Oppenheimer}, B.~R. 2010, {Direct Imaging of Exoplanets} ({Seager, S.}), 111--156

\bibitem[{{van Kooten} {et~al.}(2022){van Kooten}, {Ragland}, {Jensen-Clem}, {Xin}, {Delorme}, \& {Kent Wallace}}]{vanKooten2022}
{van Kooten}, M. A.~M., {Ragland}, S., {Jensen-Clem}, R., {et~al.} 2022, \apj, 932, 109

\bibitem[{{Vigan} {et~al.}(2011){Vigan}, {Dohlen}, \& {Mazzanti}}]{Vigan2011}
{Vigan}, A., {Dohlen}, K., \& {Mazzanti}, S. 2011, \ao, 50, 2708

\bibitem[{{Vigan} {et~al.}(2022){Vigan}, {Dohlen}, {N'Diaye}, {Cantalloube}, {Girard}, {Milli}, {Sauvage}, {Wahhaj}, {Zins}, {Beuzit}, {Caillat}, {Costille}, {Le Merrer}, {Mouillet}, \& {Tourenq}}]{Vigan2022}
{Vigan}, A., {Dohlen}, K., {N'Diaye}, M., {et~al.} 2022, \aap, 660, A140

\bibitem[{{Vigan} {et~al.}(2019){Vigan}, {N'Diaye}, {Dohlen}, {Sauvage}, {Milli}, {Zins}, {Petit}, {Wahhaj}, {Cantalloube}, {Caillat}, {Costille}, {Le Merrer}, {Carlotti}, {Beuzit}, \& {Mouillet}}]{Vigan2019}
{Vigan}, A., {N'Diaye}, M., {Dohlen}, K., {et~al.} 2019, \aap, 629, A11

\bibitem[{{Vilas} \& {Smith}(1987)}]{Vilas1987}
{Vilas}, F. \& {Smith}, B.~A. 1987, \ao, 26, 664

\bibitem[{{Wallace} {et~al.}(2011){Wallace}, {Rao}, {Jensen-Clem}, \& {Serabyn}}]{Wallace2011}
{Wallace}, J.~K., {Rao}, S., {Jensen-Clem}, R.~M., \& {Serabyn}, G. 2011, in Society of Photo-Optical Instrumentation Engineers (SPIE) Conference Series, Vol. 8126

\bibitem[{{Wilby} {et~al.}(2018){Wilby}, {Keller}, {Sauvage}, {Dohlen}, {Fusco}, {Mouillet}, \& {Beuzit}}]{Wilby2018}
{Wilby}, M.~J., {Keller}, C.~U., {Sauvage}, J.~F., {et~al.} 2018, \aap, 615, A34

\bibitem[{{Zernike}(1934)}]{Zernike1934}
{Zernike}, F. 1934, \mnras, 94, 377

\end{thebibliography}

%-------------------------------------------------------------------
\begin{appendix}
\section{Impact of the control loop features}\label{sec:appendix}
We investigate the contrast performance with the ZWFS control loop for different parameters such as the number of corrected KL modes, the loop gain, the field stop size, and the zero-point for the reference slopes. The observing conditions remain unchanged (10\,m.s$^{-1}$ wind speed, 0.7" seeing). In the nominal configuration, we assume 350 KL corrected modes, a 0.8 loop gain, and a field stop of 35\,$\lambda/D$ diameter. The calibration is set to produce a flat intensity map at the level of the ZWFS measurements. 

\subsection{Number of corrected KL modes}
\begin{figure}[!ht]
    \centering
    \includegraphics[width=\linewidth]{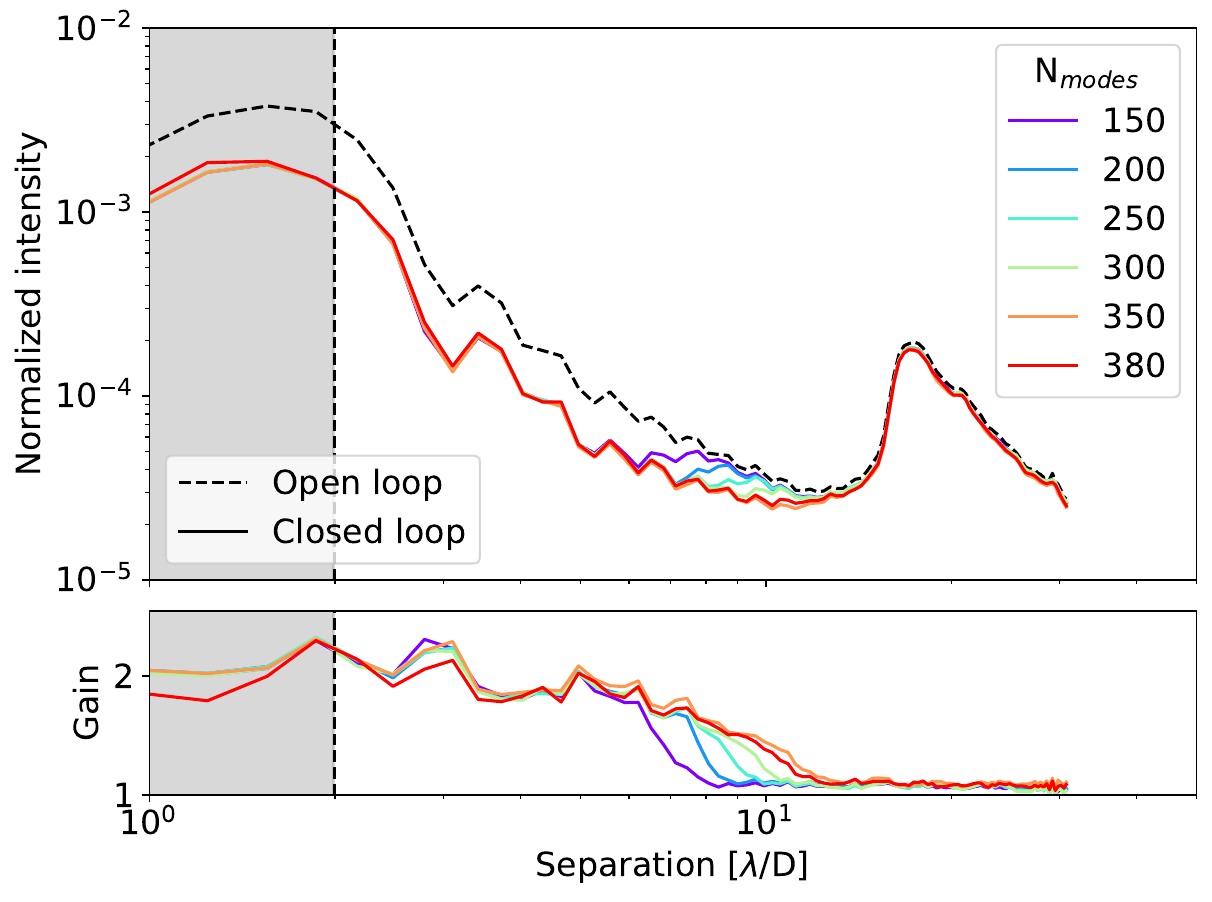}
    \caption{Contrast in the coronagraphic images with the ZWFS control loop for different numbers of controlled modes. \textbf{Top}: Normalized azimuthal averaged intensity profile of the coronagraphic images produced with the ZWFS wavefront control in open loop (dashed line) and closed loop (solid line) as a function of the angular separation for different numbers of controlled KL modes. The grey area with dashed line delimits the projected FPM size. The AO residuals are based on the VLT/SPHERE characteristics with a 6-mag natural guide star and median observing conditions with 10\,m.s$^{-1}$ windspeed and 0.7" seeing. In closed loop, the ZWFS-based second stage AO runs with an integrator using a gain of 0.8. \textbf{Bottom}: Contrast gain provided by the ZWFS-based wavefront control between the open and closed loop operations for different numbers of controlled KL modes.}
    \label{fig:cor_prf_modes}
\end{figure}

We analyze the effect of the number of corrected KL modes $N_{modes}$ on the ZWFS control loop behavior. 
Figure \ref{fig:cor_prf_modes} shows the intensity profiles of the coronagraphic images in open and closed loop for different $N_{modes}$. As expected, the ZWFS-based controlled region increases with $N_{modes}$. Beyond 350 KL modes, no further gain is observed, showing limitations in terms of wavefront correction. The DM contains 24 actuators across the pupil diameter, allowing us to control wavefront errors with spatial frequencies up to 12\,cycles/pupil (c/pup) or equivalently up to $\pi\times$12$^2$,  namely, 452 KL modes. 

Apart from the CLC performance and the NCPA, several factors can explain the contrast floor for $N_{modes}$ that are larger than 350: the limited wavefront error correction provided by the first stage XAO loop, the possible presence of residual aliasing effects, the lack of photons for the second-stage AO loop to correct for the highest-order modes. In our experiment, the somehow limited number of corrected KL modes with our control loop ($N_{modes}$=350) has no incidence on the observation of planetary companions at short separation from their host star.  

\subsection{Control loop gain}
\begin{figure}[!ht]
    \centering
    \includegraphics[width=\linewidth]{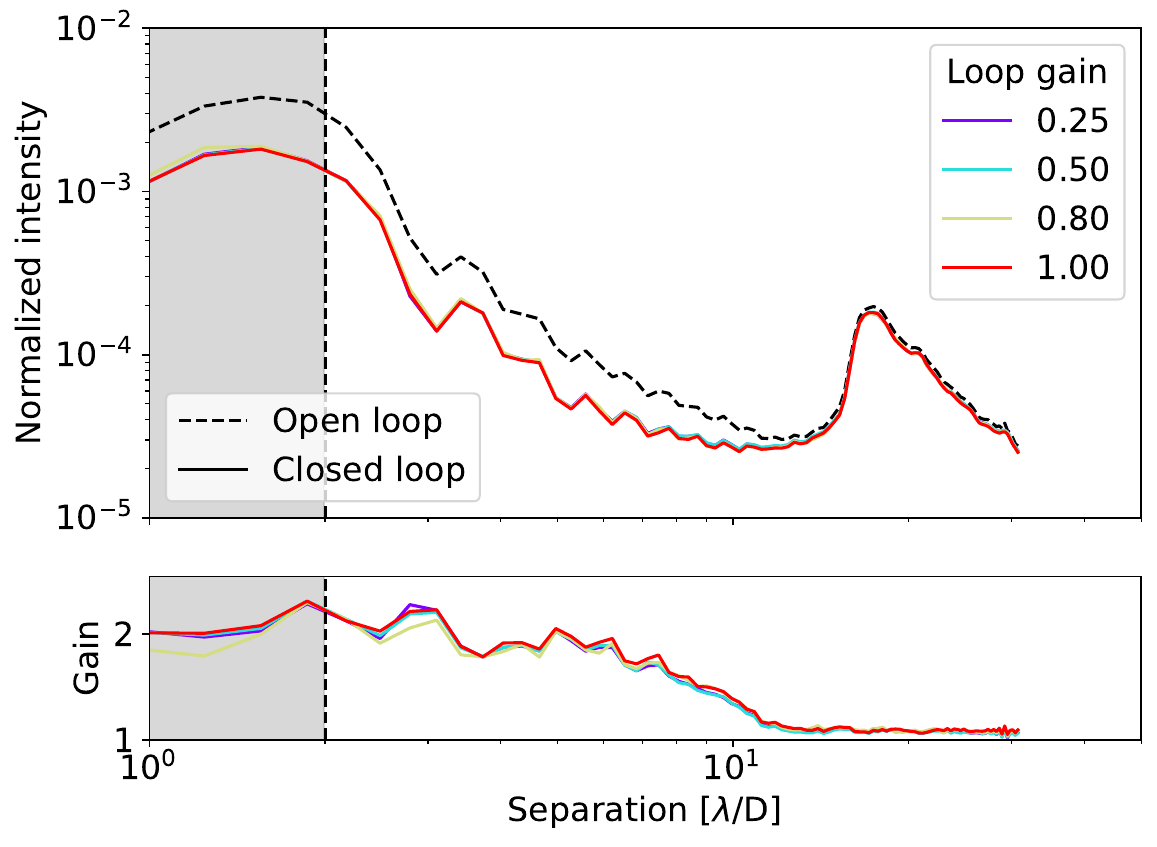}
    \caption{Contrast in the coronagraphic images with the ZWFS control loop for different integrator loop gains. \textbf{Top}: Normalized azimuthal averaged intensity profile of the coronagraphic images produced with the ZWFS wavefront control in open loop (dashed line) and closed loop (solid line) as a function of the angular separation for different loop gains. The grey area with dashed line delimits the projected FPM size. The AO residuals are based on the VLT/SPHERE characteristics with a 6-mag natural guide star and median observing conditions with 10\,m.s$^{-1}$ windspeed and 0.7" seeing. In closed loop, the ZWFS-based second-stage AO controls 350 KL modes. \textbf{Bottom}: Contrast gain provided by the ZWFS-based wavefront control between the open and closed loop operations for different loop gains.}
    \label{fig:cor_prf_gain}
\end{figure}

We investigated the impact of the control loop gain on the ZWFS-based second-stage AO performance. 
Figure \ref{fig:cor_prf_gain} displays the intensity profiles of the coronagraphic images in open and closed loops for different integrator gains. There is clearly no contrast difference between all the considered loop gains. As the ZWFS control loop runs twice faster than the first stage XAO loop, temporal error does not represent a limitation in our experiment and hence, the loop gain has no incidence in the wavefront correction and image contrast. These results are achieved with median observing conditions. In less favorable atmospheric turbulence, the loop gain remains a free parameter to adjust to allow for noisy WFS measurements to be averaged and reduce photon noise.  

\subsection{Field stop}
\begin{figure}[!ht]
    \centering
    \includegraphics[width=\linewidth]{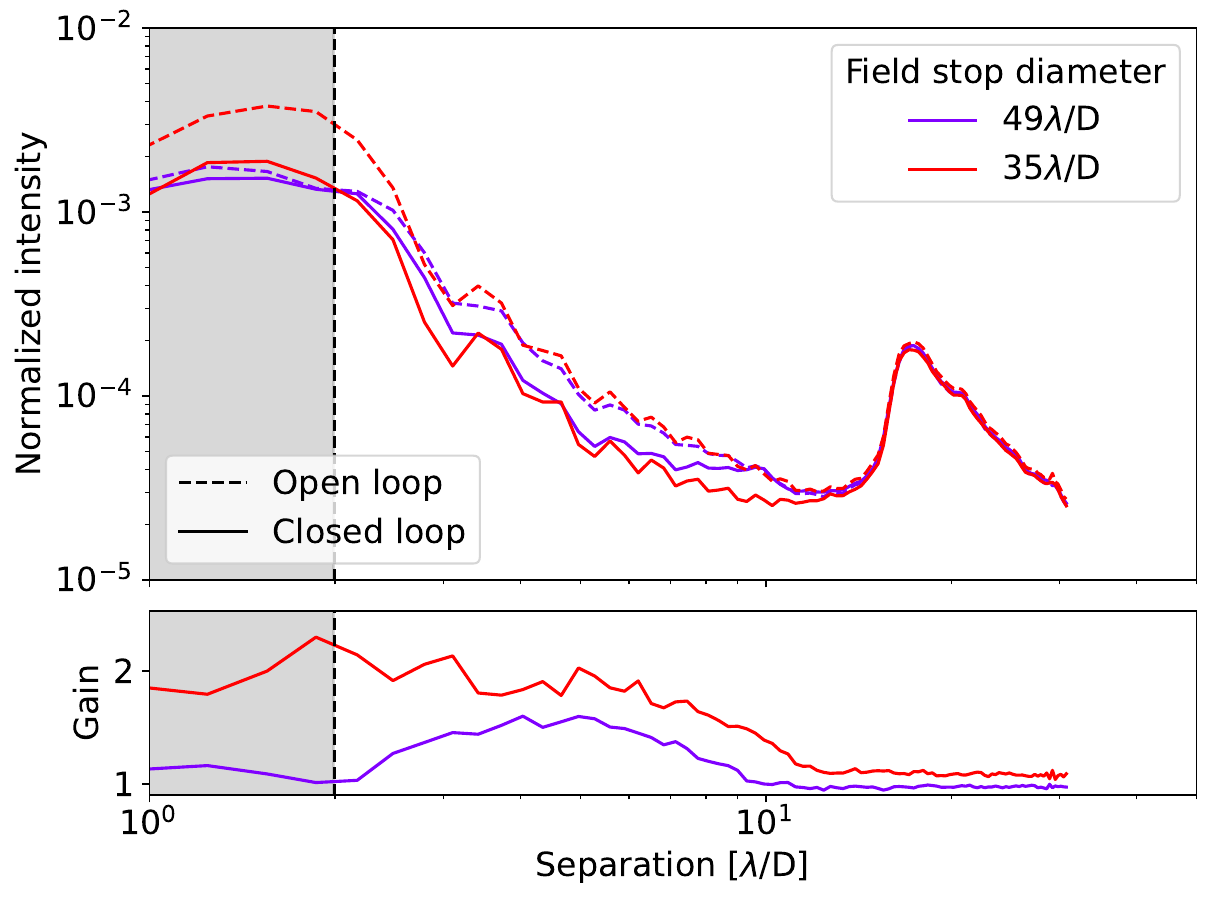}
    \caption{Contrast in the coronagraphic images with the ZWFS control loop for different field stops. \textbf{Top}: Normalized azimuthal averaged intensity profile of the coronagraphic images produced with the ZWFS wavefront control in open loop (dashed line) and closed loop (solid line) as a function of the angular separation for a field stop size of 49\,$\lambda/D$ (purple) and 35\,$\lambda/D$ (red). The grey area with dashed line delimits the projected FPM size. The AO residuals are based on the VLT/SPHERE characteristics with a 6-mag natural guide star and median observing conditions with 10\,m.s$^{-1}$ windspeed and 0.7" seeing. In closed loop, the ZWFS-based second-stage AO controls 350 KL modes and runs with an integrator using a loop gain of 0.8. \textbf{Bottom}: Contrast gain provided by the ZWFS-based wavefront control between the open and closed loop operations for different field stop sizes.}
    \label{fig:cor_prf_fieldstop}
\end{figure}

The field stop allows us to reduce the aliasing effects in the ZWFS measurements after the first XAO stage by adjusting the pinhole diameter. We study the influence of the field stop size on the behavior of our ZWFS control loop.
Figure \ref{fig:cor_prf_fieldstop} represents the intensity profiles of the coronagraphic images in open and closed loops for two field stop sizes. As expected, the reduction of the field stop size improves the control loop performance with a contrast gain going from 1.5 to 2 in median observing conditions. This result confirms the benefit of using an adjustable field stop in the second stage AO loop to reduce aliasing effects. The size of the field stop can be tuned depending on the atmospheric turbulence conditions. In the core of the paper, we keep on working with the smallest field stop with a diameter of 35\,$\lambda/D$.

\subsection{Control loop calibration}\label{subsec:calibration}
Different reference slopes can be considered for the calibration of the ZWFS control loop. By default, the reference slopes are set to correspond to a flat intensity map at the level of the ZWFS response. An alternative solution consists in performing a calibration to provide a flat map at the level of the DM, leading to a map at the level of the ZWFS which corresponds to zero aberration. We investigate the ability of our control loop to work around the DM flat map and the ZWFS flat map. 

\begin{figure}[!ht]
    \centering
    \includegraphics[width=\linewidth]{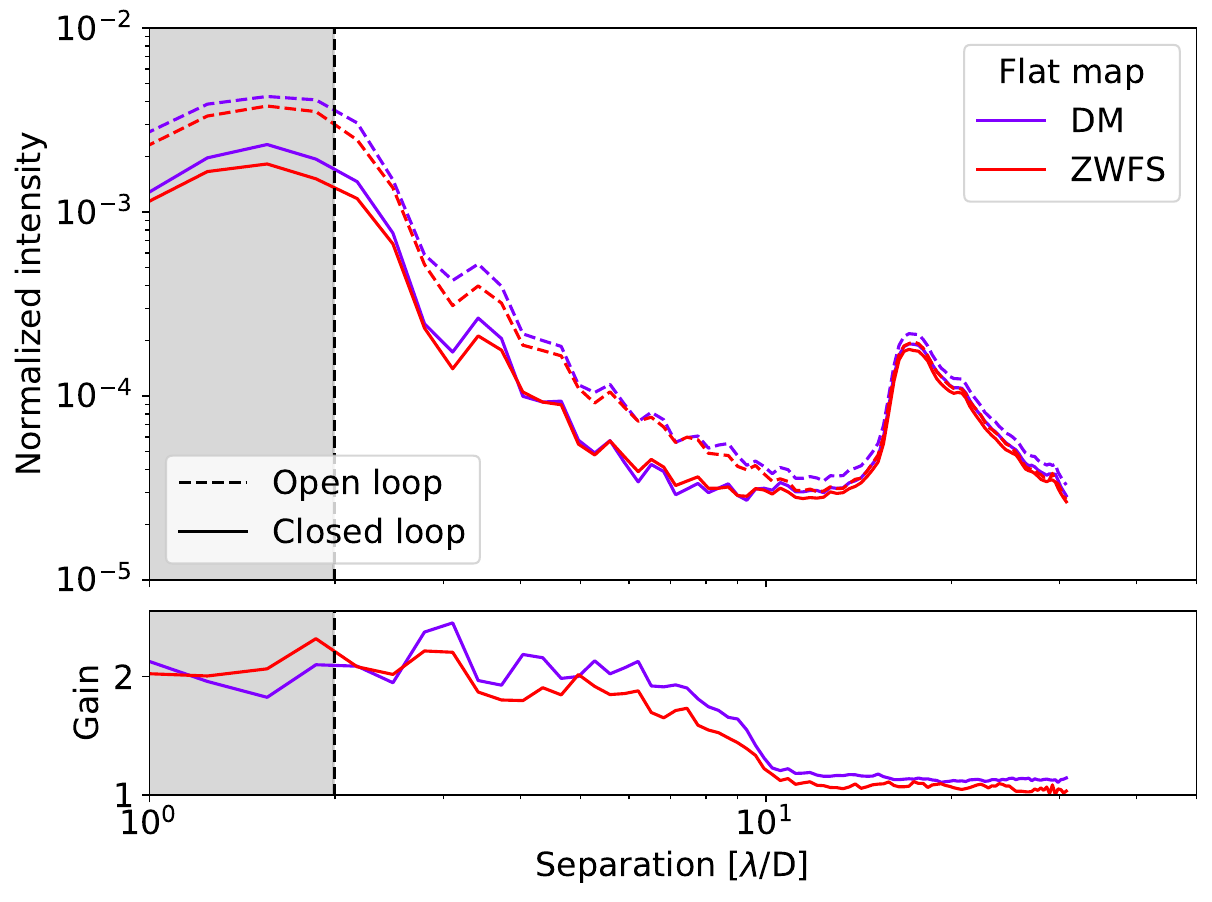}
    \caption{Contrast in the coronagraphic images with the ZWFS control loop for different flat maps. \textbf{Top}: Normalized azimuthal averaged intensity profile of the coronagraphic images produced with the ZWFS wavefront control in open loop (dashed line) and closed loop (solid line) as a function of the angular separation for the reference slopes corresponding to DM flat map (purple) and ZWFS flat map (red). The grey area with dashed line delimits the projected FPM size. The AO residuals are based on the VLT/SPHERE characteristics with a 6-mag natural guide star and median observing conditions with 10\,m.s$^{-1}$ windspeed and 0.7" seeing. In closed loop, the ZWFS-based second-stage AO controls 350 KL modes and runs with an integrator using a loop gain of 0.8. \textbf{Bottom}: Contrast gain provided by the ZWFS-based wavefront control between the open and closed loop operations for different reference slopes.}
    \label{fig:cor_prf_flatmap}
\end{figure}

Figure \ref{fig:cor_prf_flatmap} displays the intensity profiles of the coronagraphic images for both calibrations. The ZWFS control loop works in both cases with similar contrast gains, showing its effectiveness for both calibration schemes. During our experiments, our control loop proves more stable when working around the ZWFS flat map. The ZWFS is known for having an asymmetric capture range with a small linear range \citep[e.g.,][]{N'Diaye2013a}. The control loop with the ZWFS benefits from setting the reference slopes to a sensor response flat map since it provides a uniform response at which the ZWFS can properly measure wavefront errors and maintain a suitable working control loop. However, as the sensor flat response is different from the response to zero aberration as recalled in Eq.~(\ref{eq:zwfs_signal0}), it results in the introduction of a static aberration in our control loop. 

Shifting the reference slopes from the ZWFS to the DM flat map enables us to work with the system to achieve null phase errors. This comes at a cost since this calibration relies on the asymmetry in the capture range of the sensor for each subaperture. It results in higher risks when moving away from the linear regime of the sensor, leading to situations in which the control loop can break down and open. The ZWFS control loop efficiently works in different calibration situations and its performance is the most stable for the calibration around the ZWFS flat map. 

\section{Summary of the results}\label{sec:summary}
Table \ref{tab:tests_log} {summarizes} the characteristics of the observing conditions, along with the control loop parameters for all the experiments presented in the paper.

\begin{landscape}

%\Rotatebox{90}{
\begin{table}%[!ht]
    \caption{Observing conditions and the resulting contrasts for the experiments with the second-stage AO based on the ZWFS.}

    \centering
    \begin{tabular}{cccccccc|rrc|rrc|c}
    \hline\hline
        \textbf{Source} & \textbf{Seeing} & \textbf{Wind} & \textbf{Corr.} & \textbf{Loop} & \textbf{Field} & \textbf{Control} & \textbf{Flat} & \textbf{OL contrast} & \textbf{CL contrast} & \textbf{Gain} & \textbf{OL contrast} & \textbf{CL contrast} & \textbf{Gain} & \textbf{Fig.}\\
        \textbf{flux} & $\beta$ & \textbf{speed $v$} & \textbf{modes} & \textbf{gain} & \textbf{stop} & \textbf{loop} & \textbf{map} & \textbf{@5$\lambda$/D} & \textbf{@5$\lambda$/D} &  & \textbf{@10$\lambda$/D} & \textbf{@10$\lambda$/D} &  & \\
        $\Delta$mag & " & m.s$^{-1}$ & & & $\lambda/D$ & & & $\times 10^{-5}$ & $\times 10^{-5}$ & & $\times 10^{-5}$ & $\times 10^{-5}$ & & \\\hline
        &&&&&&&&&&&&&&\\

        0.0 & 10 & 0.7 & 350 & 0.80 & 35 & ZWFS & ZWFS & 12.4 $\pm$ 5.5 & 6.5 $\pm$ 4.0 & 1.9 & 3.8 $\pm$ 1.0 & 2.7 $\pm$ 0.7 & 1.4 & \ref{fig:cor_median_cond},\ref{fig:cor_prf_windspeed},\ref{fig:cor_prf_seeing},\ref{fig:cor_prf_sourceflux},\\
        &&&&&&&&&&&&&&\ref{fig:cor_prf_modes},\ref{fig:cor_prf_gain},\ref{fig:cor_prf_fieldstop},\ref{fig:cor_prf_flatmap}\\
        &&&&&&&&&&&&&&\\
        
        0.0 & 34 & 0.7 & 350 & 0.80 & 35 & ZWFS & ZWFS & 106.8 $\pm$ 42.8 & 10.9 $\pm$ 4.4 & 9.8 & 28.9 $\pm$ 11.9 & 9.0 $\pm$ 2.7 & 3.2 & \ref{fig:cor_prf_windspeed}\\
        0.0 & 24 & 0.7 & 350 & 0.80 & 35 & ZWFS & ZWFS & 46.1 $\pm$ 18.7 & 7.8 $\pm$ 3.8 & 5.9 & 12.6 $\pm$ 4.7 & 5.0 $\pm$ 1.2 & 2.5 & \ref{fig:cor_prf_windspeed}\\
%        0.0 & 10 & 0.7 & 350 & 0.80 & 35 & ZWFS & ZWFS & 12.4 $\pm$ 5.5 & 6.5 $\pm$ 4.0 & 1.9 & 3.8 $\pm$ 1.0 & 2.7 $\pm$ 0.7 & 1.4 &  \\
        0.0 &  5 & 0.7 & 350 & 0.80 & 35 & ZWFS & ZWFS & 8.5 $\pm$ 5.3 & 6.2 $\pm$ 3.8 & 1.4 & 2.7 $\pm$ 0.8 & 2.3 $\pm$ 0.6 & 1.2 & \ref{fig:cor_prf_windspeed}\\
        &&&&&&&&&&&&&&\\
        
        0.0 & 10 & 1.0 & 350 & 0.80 & 35 & ZWFS & ZWFS & 24.4 $\pm$ 8.7 & 10.3 $\pm$ 4.0 & 2.4 & 8.2 $\pm$ 1.7 & 6.3 $\pm$ 1.0 & 1.3 & \ref{fig:cor_prf_seeing}\\
%        0.0 & 10 & 0.7 & 350 & 0.80 & 35 & ZWFS & ZWFS & 12.4 $\pm$ 5.5 & 6.5 $\pm$ 4.0 & 1.9 & 3.8 $\pm$ 1.0 & 2.7 $\pm$ 0.7 & 1.4 &  \\
        0.0 & 10 & 0.5 & 350 & 0.80 & 35 & ZWFS & ZWFS & 8.8 $\pm$ 5.3 & 5.5 $\pm$ 3.8 & 1.6 & 2.5 $\pm$ 0.9 & 1.8 $\pm$ 0.7 & 1.4 & \ref{fig:cor_prf_seeing}\\
        &&&&&&&&&&&&&&\\
        
        5.0 & 10 & 0.7 & 350 & 0.80 & 35 & ZWFS & ZWFS & 12.9 $\pm$ 5.8 & 20.7 $\pm$ 4.3 & 0.6 & 4.3 $\pm$ 1.1 & 17.4 $\pm$ 2.5 & 0.2 & \ref{fig:cor_prf_sourceflux},\ref{fig:cor_prf_sourceflux_00035}\\
        4.5 & 10 & 0.7 & 350 & 0.80 & 35 & ZWFS & ZWFS & 12.8 $\pm$ 6.4 & 11.1 $\pm$ 4.3 & 1.2 & 3.8 $\pm$ 1.1 & 7.0 $\pm$ 1.1 & 0.5 & \ref{fig:cor_prf_sourceflux} \\
        2.9 & 10 & 0.7 & 350 & 0.80 & 35 & ZWFS & ZWFS & 12.8 $\pm$ 6.5 & 6.8 $\pm$ 3.9 & 1.9 & 3.7 $\pm$ 1.0 & 2.8 $\pm$ 0.8 & 1.3 &  \ref{fig:cor_prf_sourceflux}\\
        1.5 & 10 & 0.7 & 350 & 0.80 & 35 & ZWFS & ZWFS & 12.8 $\pm$ 6.4 & 6.6 $\pm$ 3.8 & 2.0 & 3.8 $\pm$ 1.0 & 2.7 $\pm$ 0.7 & 1.4 &  \ref{fig:cor_prf_sourceflux}\\
        0.7 & 10 & 0.7 & 350 & 0.80 & 35 & ZWFS & ZWFS & 12.4 $\pm$ 6.2 & 6.3 $\pm$ 3.7 & 1.9 & 3.6 $\pm$ 1.0 & 2.6 $\pm$ 0.7 & 1.4 &  \ref{fig:cor_prf_sourceflux}\\
%        0.0 & 10 & 0.7 & 350 & 0.80 & 35 & ZWFS & ZWFS & 12.4 $\pm$ 5.5 & 6.5 $\pm$ 4.0 & 1.9 & 3.8 $\pm$ 1.0 & 2.7 $\pm$ 0.7 & 1.4 &  \\
        &&&&&&&&&&&&&&\\
        
%        5.0 & 10 & 0.7 & 350 & 0.80 & 35 & ZWFS & ZWFS & 12.9 $\pm$ 5.8 & 20.7 $\pm$ 4.3 & 0.6 & 4.3 $\pm$ 1.1 & 17.4 $\pm$ 2.5 & 0.2 & \ref{fig:cor_prf_sourceflux_00035} \\
        5.0 & 10 & 0.7 & 350 & 0.15 & 35 & ZWFS & ZWFS & 13.2 $\pm$ 5.5 & 8.4 $\pm$ 3.8 & 1.6 & 4.4 $\pm$ 1.2 & 4.0 $\pm$ 1.0 & 1.1 &  \ref{fig:cor_prf_sourceflux_00035},\ref{fig:cor_prf_sourceflux_zwfs-pwfs} \\
        &&&&&&&&&&&&&&\\
        
%        5.0 & 10 & 0.7 & 350 & 0.15 & 35 & ZWFS & ZWFS & 13.2 $\pm$ 5.5 & 8.4 $\pm$ 3.8 & 1.6 & 4.4 $\pm$ 1.2 & 4.0 $\pm$ 1.0 & 1.1 & \ref{fig:cor_prf_sourceflux_zwfs-pwfs} \\
        5.0 & 10 & 0.7 & 350 & 0.03 & 35 & PWFS & PWFS & 11.2 $\pm$ 3.7 & 12.4 $\pm$ 4.9 & 0.9 & 4.3 $\pm$ 1.1 & 5.7 $\pm$ 1.3 & 0.8 & \ref{fig:cor_prf_sourceflux_zwfs-pwfs} \\
        &&&&&&&&&&&&&&\\
        
        0.0 & 10 & 0.7 & 150 & 0.80 & 35 & ZWFS & ZWFS & 12.4 $\pm$ 5.5 & 6.6 $\pm$ 4.0 & 1.9 & 3.8 $\pm$ 1.0 & 3.5 $\pm$ 0.9 & 1.1 & \ref{fig:cor_prf_modes}\\
        0.0 & 10 & 0.7 & 200 & 0.80 & 35 & ZWFS & ZWFS & 12.4 $\pm$ 5.5 & 6.5 $\pm$ 4.1 & 1.9 & 3.8 $\pm$ 1.0 & 3.4 $\pm$ 0.9 & 1.1 &  \ref{fig:cor_prf_modes}\\
        0.0 & 10 & 0.7 & 250 & 0.80 & 35 & ZWFS & ZWFS & 12.4 $\pm$ 5.5 & 6.6 $\pm$ 4.1 & 1.9 & 3.8 $\pm$ 1.0 & 3.4 $\pm$ 0.9 & 1.1 &  \ref{fig:cor_prf_modes}\\
        0.0 & 10 & 0.7 & 300 & 0.80 & 35 & ZWFS & ZWFS & 12.4 $\pm$ 5.5 & 6.6 $\pm$ 4.1 & 1.9 & 3.8 $\pm$ 1.0 & 3.0 $\pm$ 0.7 & 1.3 &  \ref{fig:cor_prf_modes}\\
%        0.0 & 10 & 0.7 & 350 & 0.80 & 35 & ZWFS & ZWFS & 12.4 $\pm$ 5.5 & 6.4 $\pm$ 4.2 & 1.9 & 3.8 $\pm$ 1.0 & 2.6 $\pm$ 0.7 & 1.5 &  \\
        0.0 & 10 & 0.7 & 380 & 0.80 & 35 & ZWFS & ZWFS & 12.4 $\pm$ 5.5 & 6.5 $\pm$ 4.0 & 1.9 & 3.8 $\pm$ 1.0 & 2.7 $\pm$ 0.7 & 1.4 &  \ref{fig:cor_prf_modes}\\
        &&&&&&&&&&&&&&\\
        
        0.0 & 10 & 0.7 & 350 & 0.25 & 35 & ZWFS & ZWFS & 12.4 $\pm$ 5.5 & 6.6 $\pm$ 4.1 & 1.9 & 3.8 $\pm$ 1.0 & 2.8 $\pm$ 0.7 & 1.4 & \ref{fig:cor_prf_gain} \\
        0.0 & 10 & 0.7 & 350 & 0.50 & 35 & ZWFS & ZWFS & 12.4 $\pm$ 5.5 & 6.6 $\pm$ 4.1 & 1.9 & 3.8 $\pm$ 1.0 & 2.8 $\pm$ 0.7 & 1.4 & \ref{fig:cor_prf_gain}\\
%        0.0 & 10 & 0.7 & 350 & 0.80 & 35 & ZWFS & ZWFS & 12.4 $\pm$ 5.5 & 6.5 $\pm$ 4.0 & 1.9 & 3.8 $\pm$ 1.0 & 2.7 $\pm$ 0.7 & 1.4 &  \\
        0.0 & 10 & 0.7 & 350 & 1.00 & 35 & ZWFS & ZWFS & 12.4 $\pm$ 5.5 & 6.5 $\pm$ 4.1 & 1.9 & 3.8 $\pm$ 1.0 & 2.7 $\pm$ 0.7 & 1.4 & \ref{fig:cor_prf_gain}\\
        &&&&&&&&&&&&&&\\
        
        0.0 & 10 & 0.7 & 350 & 0.80 & 49 & ZWFS & ZWFS & 11.3 $\pm$ 6.3 & 7.4 $\pm$ 5.5 & 1.5 & 3.9 $\pm$ 1.2 & 3.9 $\pm$ 1.3 & 1.0 & \ref{fig:cor_prf_fieldstop} \\
%        0.0 & 10 & 0.7 & 350 & 0.80 & 35 & ZWFS & ZWFS & 12.4 $\pm$ 5.5 & 6.5 $\pm$ 4.0 & 1.9 & 3.8 $\pm$ 1.0 & 2.7 $\pm$ 0.7 & 1.4 &  \\
        &&&&&&&&&&&&&&\\
        
        0.0 & 10 & 0.7 & 350 & 0.80 & 35 & ZWFS & DM & 13.6 $\pm$ 6.2 & 6.5 $\pm$ 3.0 & 2.1 & 4.1 $\pm$ 0.9 & 3.1 $\pm$ 0.7 & 1.3 & \ref{fig:cor_prf_flatmap} \\
%        0.0 & 10 & 0.7 & 350 & 0.80 & 35 & ZWFS & ZWFS & 12.4 $\pm$ 5.5 & 6.6 $\pm$ 4.1 & 1.9 & 3.8 $\pm$ 1.0 & 3.0 $\pm$ 0.7 & 1.3 &  \\\hline
        &&&&&&&&&&&&&&\\\hline
                 
    \end{tabular}
    \tablefoot{The contrasts and gain in the coronagraphic images with CLC between open and closed loop (OL and CL) are given at different angular separations. The contrast was computed in average and standard deviation in the coronagraphic image over an annular region with a width 1$\lambda$/D centered at the considered separation.}
    \label{tab:tests_log}
\end{table}

\end{landscape}

\end{appendix}

%-------------------------------------------------------------------
\end{document}